\documentclass[fleqn,usenatbib]{mnras}

\usepackage{newtxtext,newtxmath}

\usepackage[T1]{fontenc}

\DeclareRobustCommand{\VAN}[3]{#2}
\let\VANthebibliography\thebibliography
\def\thebibliography{\DeclareRobustCommand{\VAN}[3]{##3}\VANthebibliography}

\usepackage{lipsum}
\usepackage{float}
\usepackage{array}
\usepackage{hanging}
\usepackage{xurl}

\usepackage{natbib}
\usepackage{graphicx}	
\usepackage{amsmath}	

 {\begin{list}{}%
         {\setlength{\leftmargin}{#1}}%
         \item[]%
 }
 {\end{list}}
 \defcitealias{Lacy2021}{L21}
 \defcitealias{Zou2021a}{Z21}
 \defcitealias{Hatfield2022}{H22}

\title[Protoclusters in the LSST DDFs]{\textit{Spitzer}-selected $z>1.3$ protocluster candidates in the LSST Deep-Drilling Fields}

\author[H. Gully et al.]{Harry Gully,$^{1}$\thanks{E-mail: ppyhg3@nottingham.ac.uk}
Nina Hatch$^{1}$, Yannick Bah\'e$^{2,3}$, Michael Balogh$^{4,5}$, Micol Bolzonella$^{6}$, M. C. Cooper$^{7}$, \newauthor Adam Muzzin$^{8,3}$, Lucia Pozzetti$^{6}$, Gregory Rudnick$^{9}$, Benedetta Vulcani$^{10}$, Gillian Wilson$^{11}$
\\
$^{1}$School of Physics and Astronomy, University of Nottingham, Nottingham NG7 2RD, UK\\
$^{2}$Institute of Physics, Laboratory of Astrophysics, Ecole Polytechnique Fédérale de Lausanne (EPFL), Observatoire de Sauverny, 1290 Versoix, Switzerland\\
$^{3}$Leiden Observatory, Leiden University, P.O. Box 9513, 2300 RA Leiden, The Netherlands\\
$^{4}$Department of Physics and Astronomy, University of Waterloo, Waterloo, ON N2L 3G1, Canada\\
$^{5}$Waterloo Centre for Astrophysics, University of Waterloo, Waterloo, ON N2L 3G1, Canada\\
$^{6}$INAF-Osservatorio di Astrofisica e Scienza dello Spazio, Via Gobetti 93/3, 40129, Bologna, Italy\\
$^{7}$Department of Physics \& Astronomy, University of California, Irvine, 4129 Reines Hall, Irvine, CA 92697, USA\\
$^{8}$Department of Physics and Astronomy, York University, 4700, Keele Street, Toronto, ON, MJ31P3, Canada\\
$^{9}$Department of Physics and Astronomy, The University of Kansas, Malott Room 1082, 1251 Wescoe Hall Drive, Lawrence, KS 66045, USA\\
$^{10}$INAF-Osservatorio astronomico di Padova, Vicolo Osservatorio 5, I-35122 Padova, Italy\\
$^{11}$Department of Physics, University of California Merced, 5200 North Lake Road, Merced, CA 95343, USA
}

\date{Accepted XXX. Received YYY; in original form ZZZ}

\pubyear{2023}

\begin{document}
\label{firstpage}
\pagerange{\pageref{firstpage}--\pageref{lastpage}}
\maketitle

\begin{abstract}
We have identified 189 candidate $z>1.3$ protoclusters and clusters in the LSST Deep Drilling Fields. This sample will enable the measurement of the metal enrichment and star formation history of clusters during their early assembly period through the direct measurement of the rate of supernovae identified through the LSST. The protocluster sample was selected from galaxy overdensities in a \textit{Spitzer}/IRAC colour-selected sample using criteria that were optimised for protocluster purity using a realistic lightcone. Our tests reveal that $60-80\%$ of the identified candidates are likely to be genuine protoclusters or clusters, which is corroborated by a $\sim4\sigma$ stacked X-ray signal from these structures. We provide photometric redshift estimates for 47 candidates which exhibit strong peaks in the photo-$z$ distribution of their candidate members. However, the lack of a photo-$z$ peak does not mean a candidate is not genuine, since we find a stacked X-ray signal of similar significance from both the candidates that exhibit photo-$z$ peaks and those that do not. Tests on the lightcone reveal that our pursuit of a pure sample of protoclusters results in that sample being highly incomplete ($\sim4\%$) and heavily biased towards larger, richer, more massive, and more centrally concentrated protoclusters than the total protocluster population. Most ($\sim75\%$) of the selected protoclusters are likely to have a maximum collapsed halo mass of between $10^{13}-10^{14}$ M$_{\odot}$, with only $\sim25\%$ likely to be collapsed clusters above $10^{14}$ M$_{\odot}$. However, the aforementioned bias ensures our sample is $\sim50\%$ complete for structures that have already collapsed into clusters more massive than $10^{14}$ M$_{\odot}$.
\end{abstract}




\section{Introduction}
Galaxy clusters are the most massive collapsed objects in the Universe and are therefore the extreme products of the hierarchical growth of structure. Their high-redshift progenitors, protoclusters, provide insight into their formation as well as the impact these extreme environments have on galaxy evolution during the epoch of peak stellar mass growth \citep[e.g.][]{Strazzullo2013, Muldrew2018}.

Observations of protoclusters have uncovered environmentally-dependant properties such as sped-up galaxy evolution \citep{Steidel2005}, enhanced star formation \citep{Hayashi2016}, and extended Ly$\alpha$ halos \citep{Matsuda2012}. However, there are clear discrepancies in some of the relations which calls into question our understanding of protoclusters. For example, some studies find a metal enhancement in protocluster galaxies \citep{Kulas2013,Shimakawa2015} while some find a metal deficiency \citep{Valentino2015,Sattari2021}. In fact, some find no environmental dependence at all \citep{Kacprzak2015,Alcorn2019}. A metallicity enhancement or deficiency can reveal information on how the protocluster environment affects the baryon cycle of galaxies \citep{Shimakawa2015}. It is likely that these conflicting results emanate from small sample sizes (and therefore large uncertainties) but more importantly from the heterogeneity in protocluster selection.

The Deep Drilling Fields (DDFs) program of the Large Synoptic Survey Telescope \citep[LSST;][]{Ivezic2019}, provides an opportunity for innovative observations of high redshift (proto)clusters as it has a deep coverage ($AB\approx 26.2-28.7$ over 10 years) and frequent temporal sampling meaning it can identify supernovae within protoclusters. The evolving rates and relative types of supernovae within protoclusters can provide direct measurements of the chemical enrichment history, star formation and quenching history, and the stellar/supernovae feedback history that governs galaxy evolution in these environments. Measuring the rates of SNe Ia and core-collapse SNe within protoclusters can, for example, constrain IMF variations \citep[see][for a review]{Bastian2010} at the intermediate \citep{Friedmann2018} and high mass \citep{Aoyama2021} ranges, respectively, for high density environments - allowing more accurate estimates for masses and star formation histories. 

Such is the design of the LSST survey that there are expected to be tens of millions of transient events in the DDFs alone over the course of the ten years of operation. However, SNe in $z>1$ protoclusters will have their spectra shifted such that the only bands with significant flux measurements are the $z$ or $y$ bands \citep{LSST2009}, rendering colour-based redshift measurements and classifications unviable \citep[e.g.][]{Gris2023}. The flux in the $z$ and $y$ bands from a $z>1$ supernova may not be enough for any redshift estimation or classification, but it can act as a trigger for rapid spectral followup, which would be needed to classify these supernovae. The high-redshift protoclusters must therefore be located before the survey in order to pre-select the protocluster supernovae, and avoid countless contaminants.

By far the most prolific method for finding protoclusters and high redshift clusters is the \textit{Spitzer}/IRAC method devised by \cite{Papovich2008}, which locates overdensities of galaxies with red colours in the IRAC Channels 1 and 2 ($3.6\mu$m and $4.5\mu$m respectively). A colour cut of this type is able to efficiently select $z>1.3$ galaxies, regardless of galaxy age and type, by utilising the $1.6\mu$m bump. This bump is caused by a minimum in opacity of H$^-$ ions in the atmospheres of cool stars \citep[][]{John1988} which imprints itself as a maximum on the SEDs of galaxies, and does not depend on the evolutionary stage of the galaxy. At $z<1$, the $3.6\mu$m and $4.5\mu$m bands probe the stellar Rayleigh–Jeans tail, causing the [3.6]-[4.5] colours to appear blue \citep[with the exception of some dusty $z\sim0.3$ star forming galaxies; see][]{Papovich2008}. At $z\geq1$, the $1.6\mu$m bump is redshifted into the IRAC bands causing the [3.6]-[4.5] colours to appear red \citep[see e.g.][]{Simpson1999, Sawicki2002, Papovich2008}. Variations of this method have been used many times to locate clusters, such as the IRAC Shallow Cluster Survey \citep[ISCS;][]{Eisenhardt2008}, the \cite{Papovich2010} $z=1.62$ cluster, the Clusters Around Radio-Loud AGN program \citep[CARLA;][]{Wylezalek2013} and the Spitzer South Pole Telescope Deep Field survey \citep[SSDF;][]{Rettura2014} amongst others \citep[e.g.][]{Galametz2012,Muzzin2013b,Martinache2018}.

The LSST DDFs regions encompass well-studied fields: the Extended Chandra Deep Field–South (CDFS), the European Large Area Infrared Survey field South 1 (ELAIS-S1), the XMM-Large-Scale Structure Survey field (XMMLSS), and the Cosmic Evolution Survey field (COSMOS), each roughly 3.5 deg in diameter. While there is \textit{Spitzer} data available in the extended COSMOS field \citep{Annunziatella2023}, we do not include it in this study in the interest of homogeneity. Previous works have searched for clusters and their progenitors using the \textit{Spitzer}/IRAC method in these fields \citep[e.g.][]{Papovich2008}. However, in these works, the selection method was not tested and therefore the purity, completeness and bias of these protocluster samples are not known.

In this study, we address these shortcomings by making two improvements on earlier works. First, we use the deepest and most complete \textit{Spitzer} catalogues of these fields constructed by  \citet[hereafter \citetalias{Lacy2021}]{Lacy2021}. Second, we test and optimise the \textit{Spitzer}/IRAC method on a simulated lightcone in order to determine the purity, completeness and bias of the protocluster sample. 

In Section \ref{Data}, we introduce the observations and simulations we use in this work. In section \ref{Optimisation}, we use the simulations to optimise the selection method and compare the optimised selection parameters to other criteria used in the literature. Section~\ref{Sample} presents the candidate protoclusters in the DDFs, and in Section~\ref{Discussion} we compare the new catalogue to other cluster/protocluster catalogues of the field within the literature and perform X-ray stacking analysis to search for evidence of collapsed structures. Finally, we explore the biases of the protocluster sample using the simulations. Our conclusions are presented in Section~\ref{Conclusions}.

As discussed in \cite{Overzier2016}, there is no general consensus on the definition of a protocluster. One simple definition, commonly used in simulation studies, is that protoclusters are the progenitors of the massive galaxy clusters we see today -- in other words, a collection of dark matter haloes that will evolve into a virialised, $10^{14}$\,M$_{\odot}$ halo by $z = 0$. Unfortunately, such a definition is difficult to implement in a practical sense as it is almost impossible to know whether the present-day descendant of an observed structure will be a cluster or not, at least on a structure-by-structure basis. It therefore seems logical to use a more practical definition that can traverse simulations and observations. Hereafter, we refer to protoclusters as any significant galaxy overdensity (which we define quantitatively in Section~\ref{Optimisation} of this paper) on cMpc scales at high redshift ($z > 1$). For the purposes of simulations, we refer to any structure (M$_{200} < 10^{14}$\,M$_{\odot}$) that evolves into a $10^{14}$\,M$_{\odot}$ halo by $z = 0$ as a cluster progenitor. Unless stated otherwise, the halo mass definition we adopt is the mass enclosed by a sphere that has a density 200 times the critical density of the Universe (M$_{200}$). Mpc refers to proper Mpc distances, whilst cMpc refers to co-moving Mpc distances.

\section{Data}
\label{Data}

\subsection{Observations}
\label{Observations}

In preparation for LSST, \citetalias{Lacy2021} observed three of the DDFs (CDFS, ELAIS S1 and XMMLSS) with the Infrared Array Camera \citep[IRAC;][]{Fazio2004} on board the \textit{Spitzer Space Telescope} \citep[][]{Werner2004}, covering $\sim30$ deg$^2$ to a $5\sigma$ depth of $\approx2\mu$Jy (23.1 AB magnitude), in two bands centered on $3.6\mu$m and $4.5\mu$m. \citetalias{Lacy2021} produce two single-band catalogues using \textsc{SExtractor} \citep{Bertin1996}, filtered to only include sources with a signal-to-noise ratio (SNR) > 5 in the SWIRE $1.\!^{\prime\prime}9$ aperture \cite[][]{Lonsdale2003}. The dual-band catalogue was created by matching the two single-band catalogues with a $0.\!^{\prime\prime}6$ matching radius, followed by a $3\sigma$ cut for the SNR of the detection in a $1.\!^{\prime\prime}9$ radius at both $3.6\mu$m and $4.5\mu$m. The $3.6\mu$m source positions are given in the dual-band catalogue as these correspond to the smallest PSF. In this work, we use the dual-band catalogue containing 2.35 million sources, where we use the aperture corrected flux densities \citep[as per][]{Mauduit2012} in the standard SWIRE $4.\!^{\prime\prime}1$ aperture to calculate the $3.6\mu$m and $4.5\mu$m apparent magnitudes. To ensure uniformity in depth, we remove areas with a coverage of less than nine 100-second
frames in either band, which leaves a total area of 26.1 deg$^2$ across the three DDFs.

We select the high-redshift protocluster candidates solely on the basis of overdensities of red IRAC galaxies to ensure homogeneity, but further information on the candidates can be obtained from photometric redshift catalogues in these fields. For this work, we use the photometric redshift catalogue of \citet[hereafter \citetalias{Zou2021a}]{Zou2021a} based on forced photometry using a VIDEO fiducial model \citep{Zou2021b, Nyland2023}. These catalogues contain $\sim1.6$ million sources, covering 4.9 deg$^2$ and 3.4 deg$^2$ across CDFS and ELAIS S1 respectively, which corresponds to $\sim40\%$ and $\sim60\%$ of the \citetalias{Lacy2021} footprint of each field. For the XMMLSS field, we use the \citet[hereafter \citetalias{Hatfield2022}]{Hatfield2022} catalogue, which is based on the VIDEO-selected source catalogue using optical and near-infrared data from VISTA and HyperSuprimeCam. This is a hybrid photometric redshift catalogue, made using a mixture of template fitting and machine learning, that contains $\sim1.7$ million sources covering 4.7 deg$^2$ across XMMLSS (roughly $55\%$ of the \citetalias{Lacy2021} footprint). 

Galaxies in both photo-$z$ catalogues with `low quality' photometric redshift estimates are removed. For the \citetalias{Zou2021a} catalogue, this is done by making a cut of $Q_z < 1$, where $Q_z$ is the reliability parameter outputted from \texttt{EAZY} \citep{Brammer2008}. For \citetalias{Hatfield2022}, we simply use their \texttt{reliable} flag. The uncertainties in \citetalias{Hatfield2022} are significantly higher than in \citetalias{Zou2021a}, due to the different methods employed to determine the redshifts, so galaxies in \citetalias{Hatfield2022} with SNRs less than 4 are also removed. 

\subsection{Simulations}
\label{Simulations}
To optimise the \textit{Spitzer}/IRAC selection method, we use the Millennium MAMBO (Mocks with Abundance Matching in Bologna) lightcone which has an area of $3.14$\,deg$^2$ and contains $7,865,440$ galaxies with redshifts from $z=0.02$ to $z=10$ \citep[see][]{Girelli2021}. This lightcone uses the halo distribution from the Millennium dark matter $N$-body simulation \citep{Springel2005}, with the halo properties rescaled to match the \textit{Planck} cosmology\footnote{$\Omega_0=0.315$, $\Omega_{\Lambda}\,=\,0.685$, $h\,=\,0.673$, $n_s\,=\,0.961$ and $\sigma_8\,=\,0.826$ \citep[][]{Planck2014}}, using the method described in \cite{Angulo2010}. 
From the Millennium simulation, \cite{Henriques2015} built 24 lightcones deriving galaxies properties with the Munich semi-analytic model of galaxy formation. MAMBO follows a different approach, taking the sub-halo masses and their tridimensional positions of one of these lightcones to assign galaxy properties with empirical prescriptions: the stellar mass is assigned through the Stellar-to-Halo Mass relation \citep{Girelli2020} and all other properties (e.g. SFR, dust content, emission lines, gas metallicity, morphology, rest-frame and observed photometry) were attributed using the Empirical Galaxy Generator \citep[EGG;][]{Schreiber2017}. The cosmic star formation history and stellar mass functions of the lightcone agree well with observations for $z<5$. Therefore, we only use the portion of the lightcone up to $z=5$, which contains 7,218,510 galaxies ($92\%$ of the entire lightcone).

\begin{figure*}
    \centering    \includegraphics[width=\textwidth]{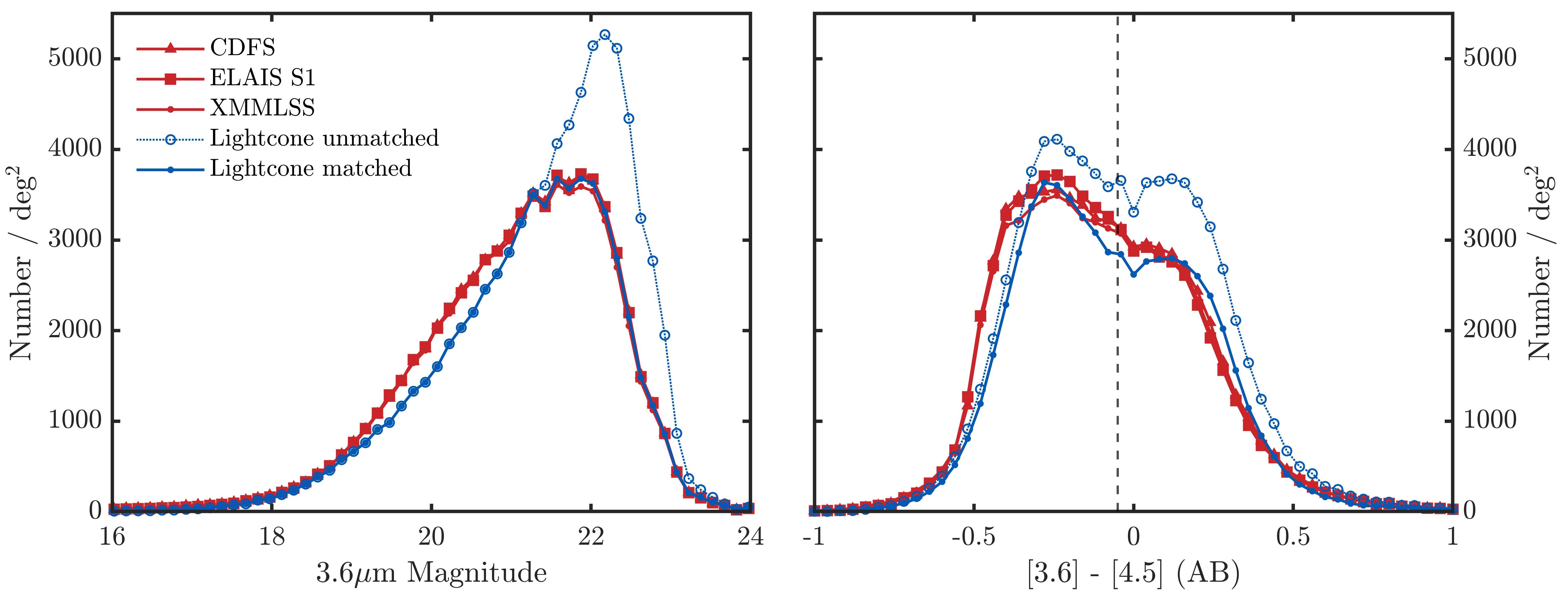}
\caption{\small{\textit{Left}: The luminosity function of all galaxies in each DeepDrill field and the lightcone with [4.5] $< 22.75$, before and after matching the source completeness in the lightcone to the DDFs. \textit{Right}: The [3.6]-[4.5] (AB) colour distribution of the same galaxies from the left panel. Black dashed line is the colour cut used in this paper ([3.6]-[4.5]$ > -0.05$).}}
        \label{Scomp}
\end{figure*}

To mimic the observational uncertainties of the \citetalias{Lacy2021} IRAC data in the lightcone, we introduce errors on the galaxy fluxes. This is done in two steps; the first is assigning each galaxy a relative error ($\delta$F/F), with the second being altering the 3.6\,$\mu$m and 4.5\,$\mu$m fluxes using those relative errors. The first step is completed by calculating the mean and standard deviation of the relative errors in the \citetalias{Lacy2021} catalogues in flux bins of width $\sim0.005\,dex$. The relative errors for the lightcone are randomly assigned assuming a Gaussian distribution using the mean and standard deviation from the real catalogues -- so as to match the relative error relationship with flux (i.e. galaxies with lower flux have larger relative errors). The $3.6\,\mu$m and $4.5\,\mu$m fluxes are then altered, assuming a Gaussian error with $1\sigma$ equal to their relative error and mean equal to their initial flux value.

We apply a magnitude cut of 22.75 (AB) in the $4.5\,\mu$m band in both the DDFs and the lightcone catalogues. However, the \citetalias{Lacy2021} catalogues only have $76\%$ completeness to this depth, so we randomly remove the appropriate fraction of  the simulated galaxies in each bin which are fainter than [3.6]$ = 21.5$ from the lightcone to ensure the galaxy number density in the simulated catalogue matches the observed catalogue. The left panel of Figure~\ref{Scomp} shows the distribution of apparent magnitudes in the $3.6\mu$m band for galaxies in the DDFs and the lightcone before and after matching the source completeness in the lightcone to the DDFs. Galaxies with [3.6]$ > 21.5$ (AB) in the lightcone were randomly removed until the number density in each magnitude bin matched the mean number density of the DDFs in the equivalent bin. Each time we perform the \textit{Spitzer}/IRAC method on the lightcone, we use a different realisation of this random removal of galaxies. We match in [3.6] as this gives a better match for the colour distribution than if we matched in [4.5]. We can also see that the lightcone under predicts the number of brighter galaxies (i.e. [3.6]$ < 21.5$ AB), however this mostly translates into an under prediction of blue IRAC galaxies so it does not affect the red IRAC galaxies that are the focus of our study . The right panel of Figure~\ref{Scomp} shows the IRAC colour distribution of galaxies in the DDFs and the lightcone before and after accounting for the higher completeness of faint galaxies. It shows that the abundance of red galaxies in the lightcone matches the DDFs well, but the number of blue galaxies in the lightcone is underestimated. As these missing bright galaxies are blue, they are likely to be at $z < 1$ and so will not have a significant effect on our study after performing the red IRAC cut.

\section{Optimising the IRAC protocluster detection method}
\label{Optimisation}

We search for protoclusters as true overdensities of galaxies in the physical coordinates of the lightcone, irrespective of whether they end up as clusters by $z=0$. We calculated the local density ($D_{gal}$) of each galaxy as the number of neighbouring galaxies within a spherical volume with a radius of $2.5$ cMpc. These values were broadly matched to the size of the density peaks in Hyperion, which is a collection of $z\sim2.4$ protoclusters in the COSMOS field \citep[coined a proto-supercluster;][]{Cucciati2018}, in order to optimise our detection algorithm for these types of objects. The overdensities ($\delta_{gal}$) were calculated with respect to the the mean density in a line-of-sight slice, $\langle D_{gal} \rangle$, of width $20$ cMpc\footnote{This width is chosen so that we can explore structures on these scales, which protoclusters typically are \citep{Muldrew2015,Lovell2018}.}, where $\delta_{gal} = (D_{gal} - \langle D_{gal} \rangle) / \langle D_{gal} \rangle$. In order to determine what overdensity selection will identify protoclusters that are cluster progenitors, we calculate the purity of selected galaxies and completeness of the selected overdensities with respect to the cluster progenitors in the lightcone (see how cluster progenitors are located in Appendix~\ref{Appendix}). We choose the overdensity value at the crossover point of purity and completeness as a compromise between the two, giving our 3D overdensity selection of $\delta_{gal}=2.63$. This corresponds to a purity and completeness of cluster progenitor galaxies of $\sim75\%$. 

\begin{figure}
    \centering
    \includegraphics[width=\columnwidth]{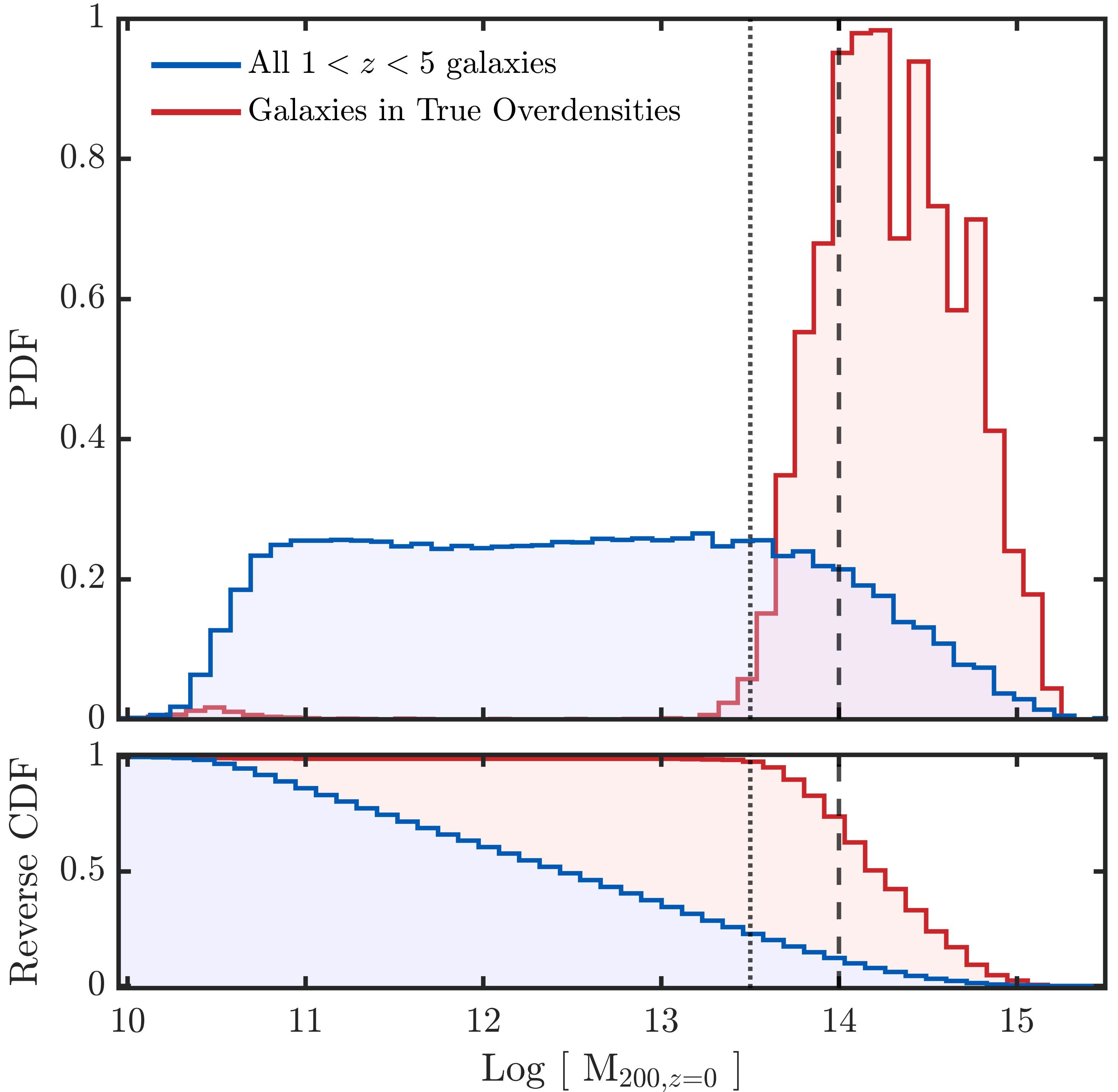}
    \caption{\small{\textit{Top}: Distribution of $z=0$ halo masses for all galaxies with $1<z<5$, and galaxies in true overdensities in the lightcone. \textit{Bottom}: Reverse cumulative distribution of galaxies as above. $74.5\%$ of selected galaxies end up in halos with mass $M /$ M$_{\odot} \geq 10^{14}$ by $z=0$ (dashed line) and $98.6\%$ end up in halos with $M /$ M$_{\odot} \geq 10^{13.5}$ by $z=0$ (dotted line).}}
        \label{WhatareTOs}
\end{figure}

The selected galaxies in overdense regions are grouped together using the density-based clustering algorithm DBSCAN \citep{Ester1996}, which was chosen as it does not require specifying the number of groups in advance and is effective in discovering groups of arbitrary shapes. DBSCAN works by identifying core points with a minimum number of neighbors within a specified radius, then expanding clusters by connecting reachable points. This results in the identification of 1,769 unique protoclusters (with at least 15 members) from $1 < z < 5$, containing 122,303 protocluster galaxies, of which $74.5\%$ end up in clusters by $z=0$. In fact, in Figure \ref{WhatareTOs}, we can see that $98.6\%$ of the galaxies in these true overdensities end up in halos of mass $M \geq 10^{13.5}$ M$_{\odot}$ at $z = 0$. Therefore, we can be satisfied that our selection of true overdensities is accurately identifying the cores of the progenitors of high mass groups and clusters. 

Having identified the true protoclusters in the lightcones we now use the MAMBO simulations of the IRAC fluxes of protocluster galaxies to determine the optimal IRAC colour and aperture to select protoclusters. The optimal parameters depend on whether the completeness or purity of the protocluster sample is considered most important. The goal for our protocluster sample is to measure the supernova rate in protoclusters, hence we must locate as pure a sample of protoclusters as possible whilst a high level of completeness is not a priority. This is because follow-up of the high-redshift supernovae is expensive and we must concentrate on only the most likely candidates. We therefore chose to optimise purity and we quantify the bias of this highly incomplete sample in Section~\ref{Discussion}. 

We measure the number density, D(r $<R$), of red IRAC galaxies within an aperture of radius $R$, centred on each of the red IRAC galaxies. To measure the reference field density, $\langle$D$\rangle$,  and its standard deviation, $\sigma_D$, we follow the method of e.g. \cite{Papovich2008} and \cite{Wylezalek2013} and fit a Gaussian to this distribution, iteratively clipping at $2\sigma_D$ to not bias our field value by overdense outliers (i.e. a fit to the low-density half of the distribution). Finally, we measure and select overdense galaxies (and their surrounding galaxies), using (D(r $<R$) $-$ $\langle$D$\rangle$) / $\sigma_D$. These galaxies are then grouped using DBSCAN. We apply this method both the the simulations and to the observational data.

We define the purity as the ratio of `successful' protocluster selections to the total number of protoclusters selected. We consider a successful protocluster selection as one in which a significant fraction of galaxy members belong to a single protocluster. We choose this fraction to be the proportion of protocluster galaxies within our mock DDF ($\sim7\%$). This is chosen as it tells us whether a group has a higher fraction of protocluster galaxies than the average within the field. Although this value seems quite low, as we show in Figure~\ref{WhatareTOs} this allows us to securely select overdensities that become group and cluster-mass objects by $z=0$. We calculate the errors on the purity by combining in quadrature the standard deviation of the purity over 100 runs of the method with the standard error of a proportion\footnote{$\sigma_p = \sqrt{\bar{P}(1-\bar{P})/\bar{N}}$, where $\bar{P}$ is the mean purity and $\bar{N}$ is the mean number of groups selected. This assumes the normal approximation to the binomial holds.} on the mean purity of 100 runs of the method. We do this to be more conservative with the accuracy of our measurements. 

\begin{figure}   \includegraphics[width=\columnwidth]{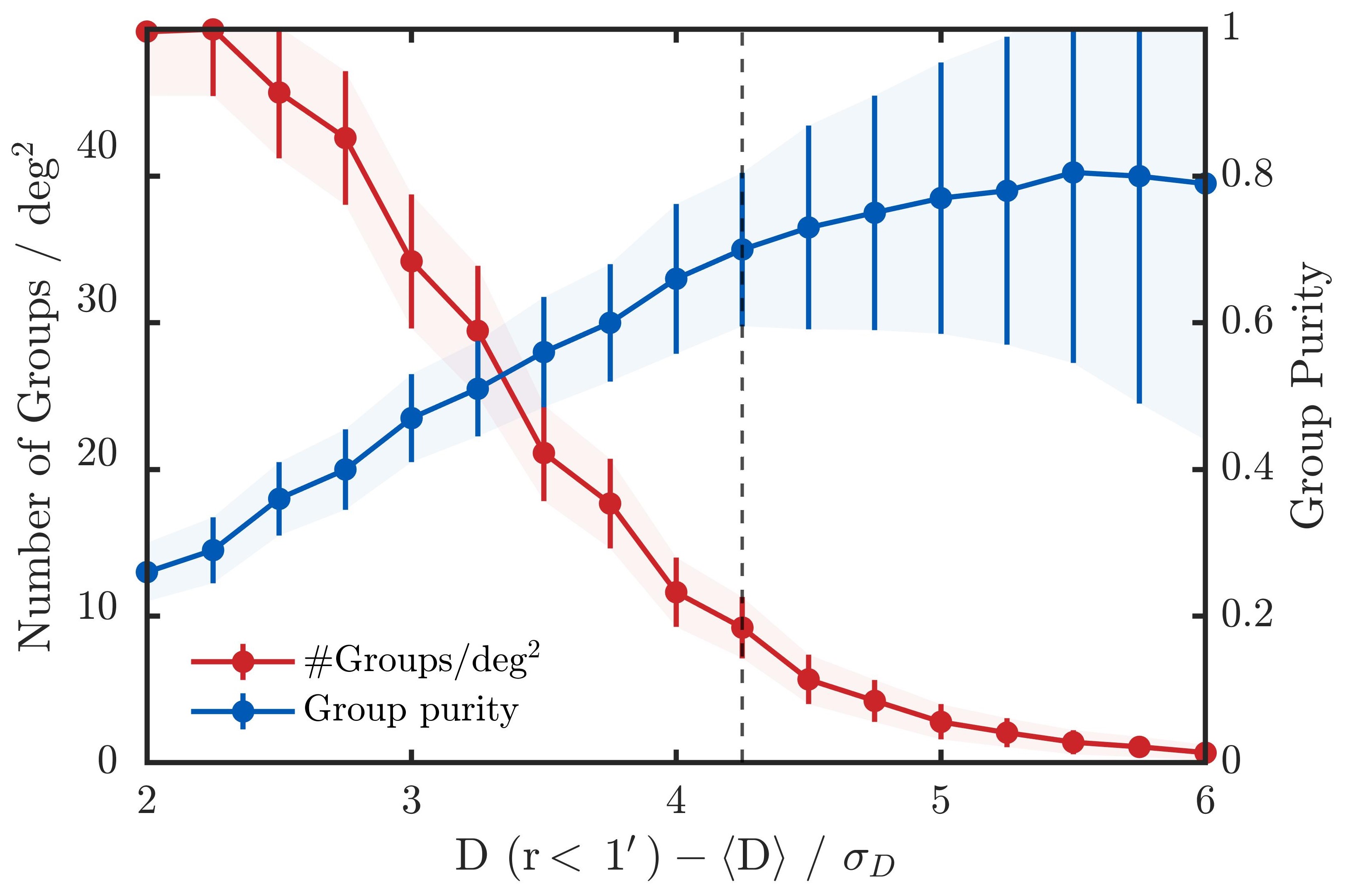}
    \caption{\small{The group purity, which is defined as the number of `successful' selections (defined in the text) over the total number of groups selected, as a function of overdensity threshold (blue), after using a colour cut of [3.6] - [4.5] $> -0.05$ and search radius of 1 arcminute. The number density of groups selected as a function of overdensity is shown in red. The overdensity threshold we use ($4.25\sigma_D$) is shown by the dashed black line. See text for error calculation.}}
        \label{PurComp}
\end{figure}

Figure~\ref{PurComp} shows how this purity varies with the overdensity selection, where we can see that higher overdensity selections produce a purer sample compared to lower selections which are more contaminated. The figure also shows how the number density of groups selected varies with overdensity selection, where we can see that higher overdensity selections yield fewer detections (we show number density as opposed to completeness for visualisation purposes as we explore completeness in Section~\ref{Biases}). 

\begin{figure}    \includegraphics[width=\columnwidth]{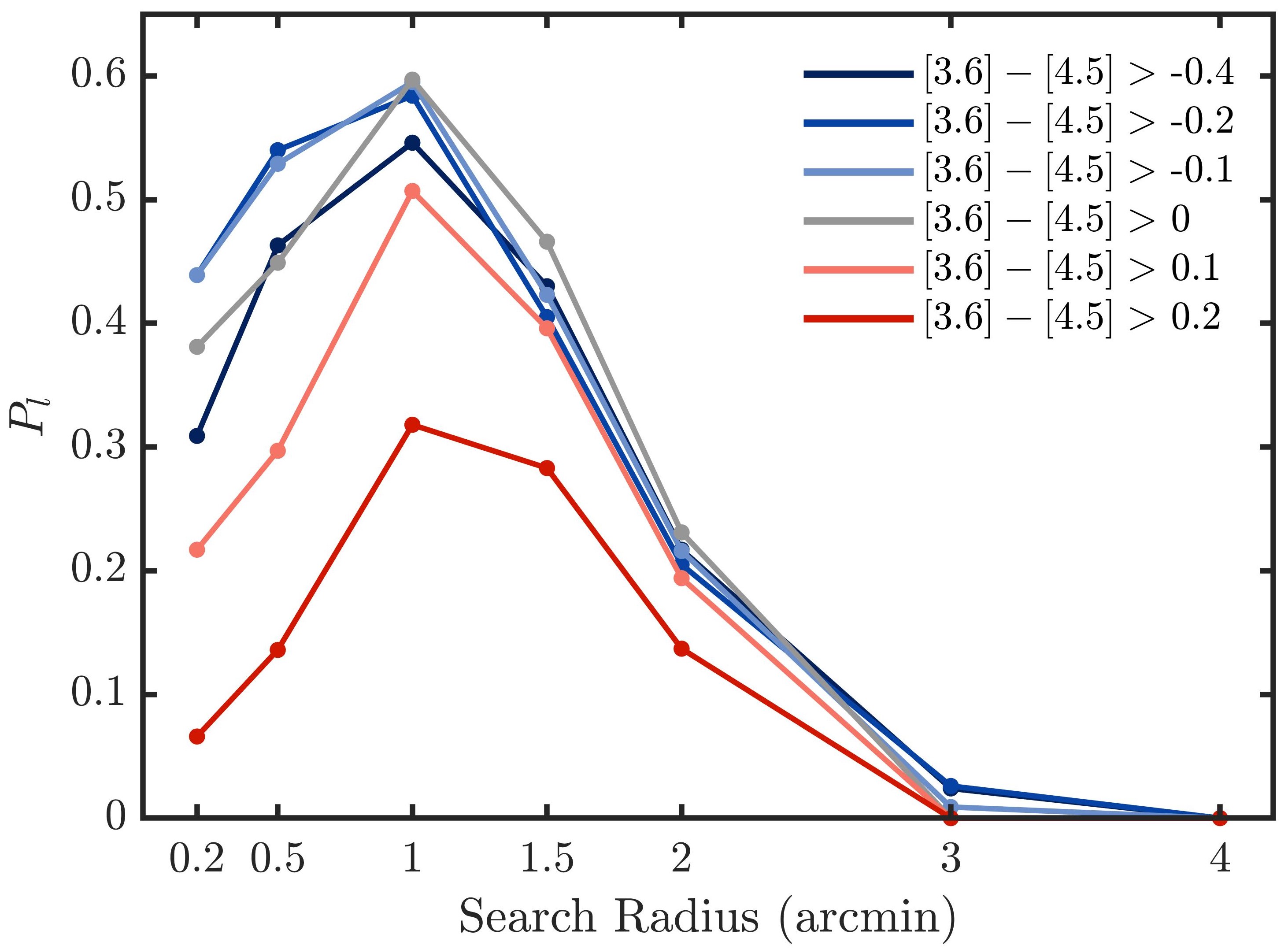}
    \caption{\small{The maximum lower bound on group purity shown as a function of the search radius used to calculate local densities. Different coloured curves represent varying colour cuts used.}}
        \label{Opt}
\end{figure}

We optimise for the maximal lower bound on group purity ($P_l$; \textit{i.e.} the lower error bar of the blue curve in Figure~\ref{PurComp}), by varying the value of the [3.6]-[4.5] colour cut, and the radius of the search area. Figure~\ref{Opt} shows how $P_l$ varies for different combinations of colour cut and search radius. Extreme red cuts ([3.6]-[4.5]$ > 0.1$) essentially select AGN \citep[see][]{Stern2012}, so they remove the majority of non-active high redshift galaxies which results in decreasing the purity of the protocluster sample. We also find that an extreme blue cut ([3.6]-[4.5]$ > -0.4$) is also not optimal as there are too many low redshift contaminants. However, we do find that in the range [3.6]-[4.5]$ > [-0.2,0]$, $P_l$ varies little  (for radii $ \geq 1^{\prime}$). The colour cut presented in \cite{Papovich2008}, [3.6]-[4.5]$>-0.1$, is the most commonly used cut in the literature \citep[e.g.][]{Galametz2012,Wylezalek2013,Rettura2014,Martinache2018}. This falls in our optimal range, but we instead opt for a value of [3.6]-[4.5]$>-0.05$ as this gives the closest match in field densities of the lightcone and DDFs (not shown) - giving us the most precise comparison to perform our tests on. 

Figure \ref{Opt} also shows that search radii $r>1.5^{\prime}$ perform particularly badly at identifying protocluster. These larger radii have a much higher probability of including chance line-of-sight alignments (scaling with $\propto r^2$), and require a substantially greater number of galaxies to yield significant overdensities. Ultimately, this results in a lower purity. However, search radii that are too small (i.e. $r<0.5^{\prime}$) also do not perform that well. While there is less likely to be a chance line-of-sight alignment, smaller radii are actually more sensitive to them (as well as noise), which can result in artificial density enhancements and false detections. We find the optimal range for search radii as $0.5^{\prime}<r<1.5^{\prime}$, and so opt for a value of $1^{\prime}$. We also checked how the magnitude limit effects the purity but found that it makes little to no difference for [4.5] $> 22-23$ mag . Using a colour cut of [3.6]-[4.5]$>-0.05$ and search radius of $r=1^{\prime}$, the highest value of $P_l$ occurs when we make our selection at an overdensity of $4.25\sigma_D$ (see Figure~\ref{PurComp}), giving a purity of protocluster detections of $70 \pm 11\%$. In Section~\ref{Biases}, we explore the biases of this sample using the lightcone, and show that it is less than $5\%$ complete and biased to the most massive halos.

\subsection{Comparisons to selection criteria used in the literature}
While we have settled on these optimal values of the parameters, other studies involving similar methods have used different values. We therefore test how values used throughout the literature perform when applied to the lightcone. \cite{Papovich2008} search for overdensities of high redshift galaxies over 50 deg$^2$, using a colour cut of [3.6] - [4.5] $> -0.1$, a search radius of $r = 1.4^{\prime}$, and an overdensity selection of $3\sigma_D$. They cover the same fields as in our work, however they use data from the SWIRE legacy survey, which only reaches a $5\sigma$ flux limit of $5.4\,\mu$Jy, which is further reduced in practice to 7-10 $\mu$Jy (21.79 - 21.4 mag) after applying S/N requirements. Using these values on the lightcone, we report a purity of only $38 \pm 9\%$ - motivating our reapplication of the \textit{Spitzer}/IRAC method on these fields. 

\cite{Rettura2014} present 279 galaxy cluster candidates over 94 deg$^2$ in the Spitzer South Pole Telescope Deep Field survey, identified as overdensities of high redshift galaxies using a [3.6] - [4.5] $> -0.1$ colour cut, a $1^{\prime}$ search radius and a $5.2\sigma_D$ overdensity selection. They also include an additional requirement on the 4.5$\mu$m band of $19.5 < [4.5]$ as well as a non-detection requirement in the SuperCOSMOS \textit{I}-band data ($I > 20.45$). With a flux limit of $9.4\,\mu$Jy in the 4.5$\mu$m band, the upper magnitude limit is $[4.5] < 21.46$. We must note that their method differs slightly as they measure overdensities with respect to local regions as opposed to the field as a whole, and they make completeness corrections that we do not -- but we do not believe this would significantly affect the results. Using these parameter values on the lightcone, we measure a purity of $57 \pm 25\%$. Interestingly, if we remove the $19.5 < [4.5]$ requirement, the purity measurement becomes $73 \pm 19\%$, and removing the $I$-band cut has relatively little effect. These cuts were introduced to reduce the number of low redshift contaminants left over from the IRAC cut but they actually worsen the purity of the final sample. While these cuts do result in a $25\%$ decrease of $z < 1$ galaxies (decreasing the contamination), there is also a $10\%$ decrease of $1 < z < 2$ galaxies. This reduction in $z>1$ galaxies results in fewer true protoclusters exhibiting a significant galaxy overdensity, which overall decreases the effectiveness of the protocluster detection method.

\cite{Martinache2018} and \cite{Wylezalek2013} are two examples of using the \textit{Spitzer}/IRAC method around high redshift targets to identify protoclusters. \cite{Martinache2018} search around bright, highly star forming galaxies and \cite{Wylezalek2013} search around high-redshift radio galaxies (H$z$RGs). These targets are thought to trace protoclusters in the early Universe, where they are found to preferentially lie in high density regions \citep[see also][]{Galametz2012,Hatch2014}. Such searches are therefore more efficient in locating protoclusters. It is beyond the scope of this paper to test the potential biases of these searches, but we can test the sample purity. \cite{Martinache2018} make a magnitude cut at $[4.5] < 22.9$, a colour cut of $[3.6] - [4.5] > -0.1$, and use a search radius of $1^{\prime}$ to identify overdensities. They find that $46\%$ of the fields around their targets have at least one 3$\sigma_D$ overdensities and $15\%$ have 4$\sigma_D$ overdensities. Applying the method on the lightcone using these parameters, we find a purity of $46 \pm 6\%$ for the 3$\sigma_D$ overdensities, and $67 \pm 11\%$ for 4$\sigma_D$ overdensities. \cite{Wylezalek2013} use the same parameter values as \cite{Martinache2018}, except identify their overdensities at a $2\sigma_D$ level. At this level, we predict only $27 \pm 5\%$ of the selected structures will be successful detections.

We caveat the above analysis with the fact that there are differences between the way we have constructed the mock catalogue in the lightcone, and the way each of the aforementioned studies construct their catalogues. Therefore, none of the purity measurements relating to these studies are to be taken as exact. However, the trends we find are robust, such as an extreme decrease in purity for studies using a low overdensity threshold ($< 4\sigma_D$), with a similar purity decrease (though far less extreme) for studies using shallower data.

One other variation of the \textit{Spitzer}/IRAC method used in the literature is the Stellar Bump Sequence (SBS) method developed by \cite{Muzzin2013b}. Instead of the single mid-infrared (MIR) $3.6\mu$m - $4.5\mu$m colour cut, they also introduce an optical/MIR $z^{\prime}$ - $3.6\mu$m colour cut in order to eliminate foreground ($0.2 < z < 0.4$) galaxies. Unfortunately, there is no $z^{\prime}$-band data available covering the entire DDFs that is deep enough to be able to incorporate into the method we use. However, we can still test its effect using the lightcone, for reference when $z^{\prime}$-band data does become available (which it will with LSST). We do not follow the exact method of searching for overdensities of galaxies in MIR colour slices, as presented in \cite{Muzzin2013b}, as we are only interested in structures at $z>1.3$ where the MIR colour is approximately constant. To form the clearest comparison possible, we use a magnitude cut of [4.5] $ < 22.75$, MIR colour cut of [3.6] - [4.5] $ > -0.05$ and search radius of $1^{\prime}$ (our optimal values), with the only difference being the introduction of the optical/MIR colour cut of $z^{\prime}$ - $3.6\mu$m $ > 1.7$. Using this criteria on the lightcone, we find the purity at the maximum value of $P_l$ is $82\pm17\%$, at an overdensity threshold of $5\sigma_D$. This result suggests that the introduction of an optical/MIR colour cut increases the purity of the selected sample, and so should be incorporated into the detection method of protoclusters in the DDFs when the data becomes available with LSST. 

\section{A sample of protocluster candidates in the DDFs}
\label{Sample}

We apply our optimal selection criteria ($1^{\prime}$ search radius, [3.6]-[4.5]$>-0.05$ colour cut, $4.25\sigma_D$ overdensity cut) to the \citetalias{Lacy2021} catalogues and we find 189 candidate protoclusters containing 15,856 red IRAC galaxies. Out of the 189 candidate protoclusters, we expect $\sim 70\%$ (113 to 151) to be true detections based on our lightcone tests. The positions of these selected structures are given in Table~\ref{TabGalsE} (available online) and are calculated as the centroid of the constituent IRAC galaxies. To determine the accuracy of these positions, we calculate the offset between our identified structures and the true protoclusters (which are calculated as the centroid of member galaxies) in the lightcone. The median distance is $40^{\prime\prime}$, with the 5th - 95th percentile range being $12^{\prime\prime} - 130^{\prime\prime}$. Therefore, the positions are off by at most $\sim2^{\prime}$.

\begin{figure*}
    \centering
    \includegraphics[width=\textwidth]{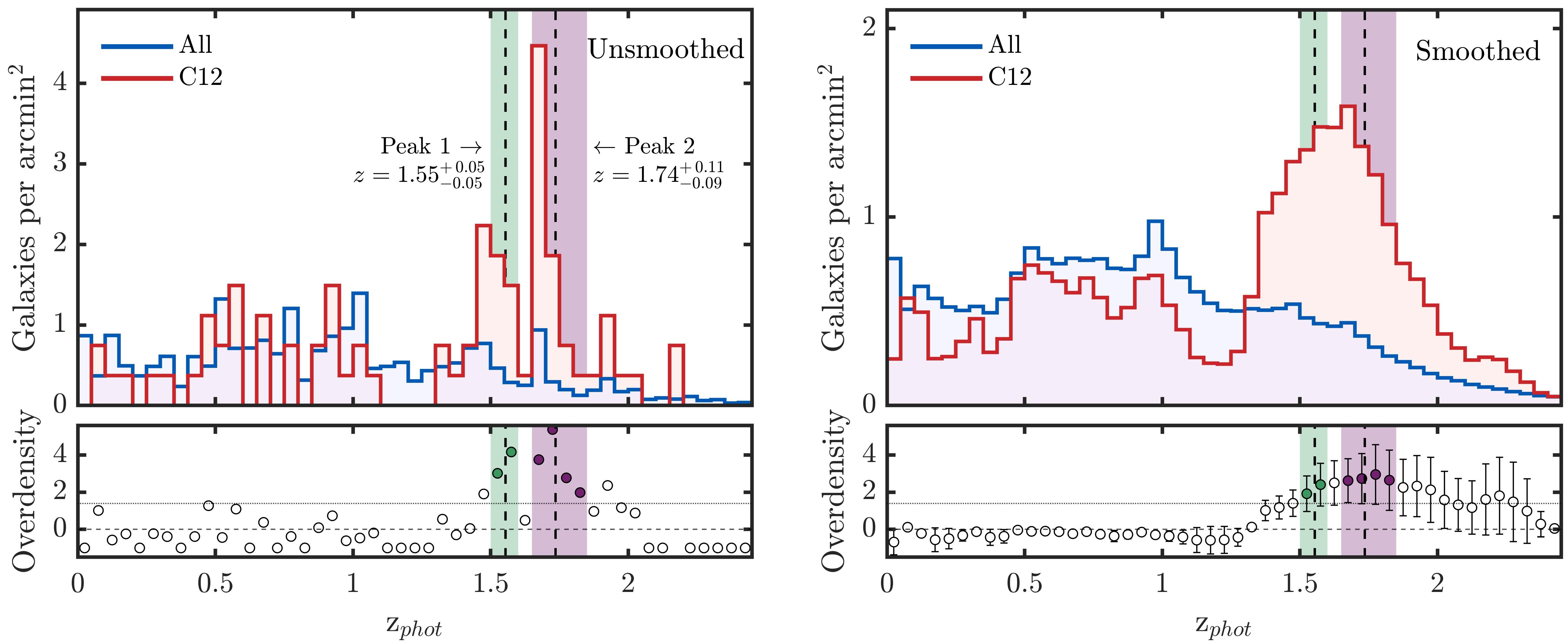}
    \caption{\small{\textit{Left}: Redshift distribution of all galaxies in the CDFS (blue) and those that are found within the projected boundary of the example group C12 (red). \textit{Right}: Same as left panel except averaged over 1,000 realisations of varying the redshifts within their errors. The bottom panels on both sides are the residuals or overdensities, with the photometric redshift overdensity peaks highlighted in green and purple.}}
        \label{PhotDist}
\end{figure*}

We use the photometric redshift catalogues of \citetalias{Zou2021a} and \citetalias{Hatfield2022} to estimate the redshifts of our candidate protoclusters for those in the overlapping area.  The photometric redshift distribution of the clean sample of galaxies in CDFS is shown in blue in the left panel of Figure~\ref{PhotDist}, with the distribution of those that fall within the boundary of example protocluster candidate C12 shown in red. Identifying redshift peaks from these distributions is possible, however it does not take into account the errors on the photometric redshift estimates. For this reason, we also produce a smoothed redshift distribution where errors are accounted for. \citetalias{Zou2021a} provide lower and upper $68\%$ confidence limits for the redshift of each source, whereas \citetalias{Hatfield2022} provide a single $68\%$ confidence limit. Therefore, for sources in ELAIS S1 and CDFS, we fit a half Gaussian below and above the given redshift value with standard deviation equal to the lower and upper bound respectively. For sources in XMMLSS, we fit a single Gaussian with standard deviation equal to the confidence limit and mean equal to the given redshift value. We bin the redshifts as before, except with values sampled randomly from our fitted Gaussians, giving a slightly different distribution each time. This is performed 1,000 times and averaged, giving the smoothed distributions in the right hand plot of Figure~\ref{PhotDist}.

The photometric redshift overdensities are calculated as $(Z_P - Z_A) / Z_A$, where $Z_P$ is the redshift distribution of all sources that are within the projected conforming boundary of the candidate protocluster, and $Z_A$ is the redshift distribution of all sources within the given field. This is done for both the unsmoothed and smoothed redshift distributions in redshift bins of $\Delta z = 0.1$. These photometric redshift overdensities are shown in the bottom panels of Figure~\ref{PhotDist} (the errors in the bottom right panel are calculated by propagating the $1\sigma$ uncertainties from $Z_P$ and $Z_A$ of the smoothed distributions). 

A redshift peak is identified if the overdensity value in the given bin is greater than 1.4 in both the unsmoothed and smoothed distributions (with an extra requirement that the lower bound in the overdensity of the smoothed distribution is greater than 0.75). These choices are arbitrary and have been chosen to match visual inspections. The overdense bins for the group C12 are shown as filled circles in the bottom panels of Figure~\ref{PhotDist}. Consecutive overdense bins are classed as the same redshift peak, with the redshift estimate (dashed black lines) calculated by an overdensity weighted average on the redshifts of the overdense bins in the unsmoothed distribution. The number of galaxies that fall within each redshift peak (shown as the shaded regions in Figure~\ref{PhotDist}) are given in Table~\ref{TabGalsE}, as well as the weighted average and width of the peak. It must be noted that there are fewer protoclusters with redshift peak estimates in XMMLSS than the other two fields due to the higher redshift uncertainties within \citetalias{Hatfield2022}. 

\begin{figure}   \includegraphics[width=\columnwidth]{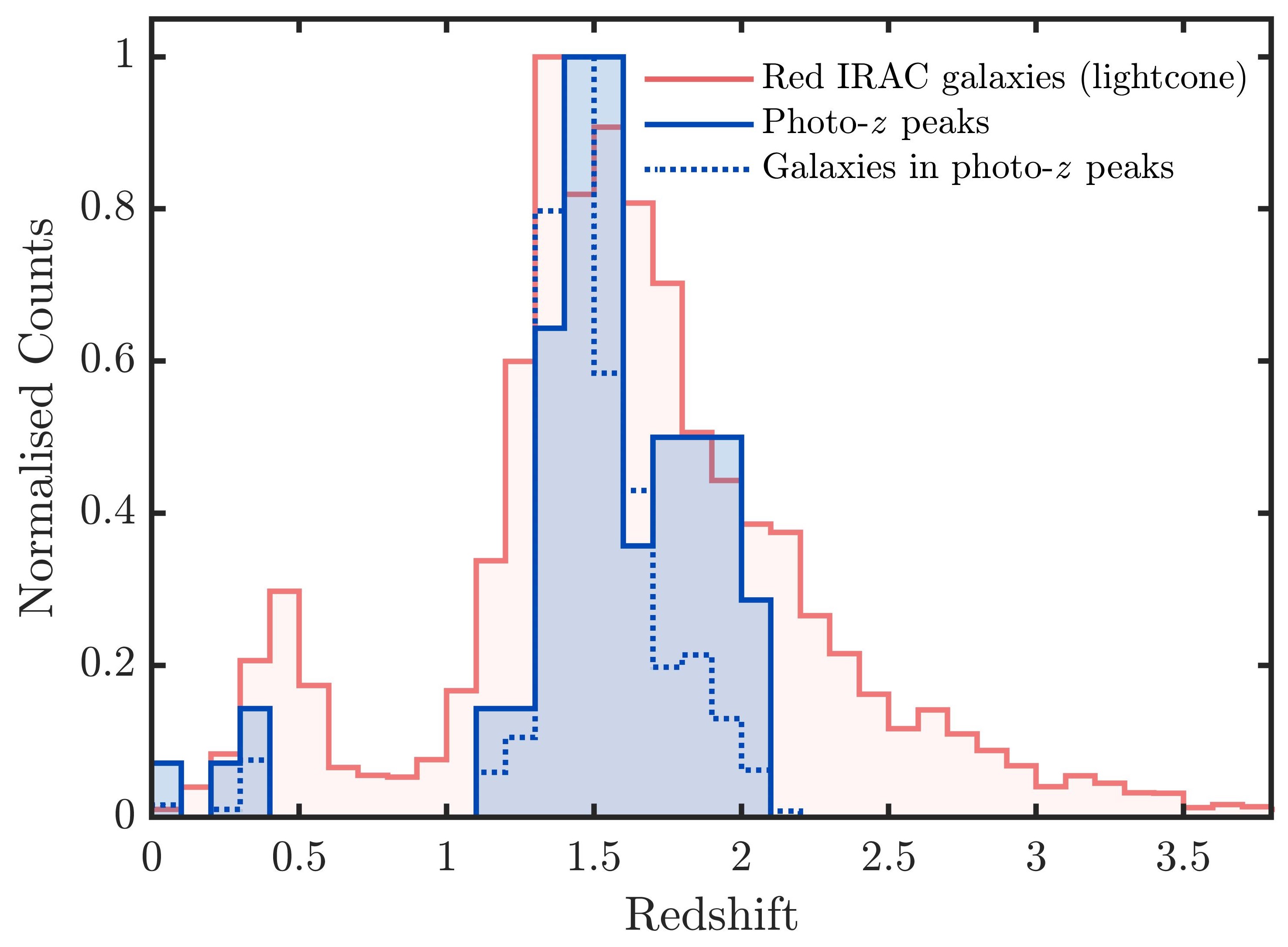}
    \caption{\small{The redshift distribution of galaxies that fall within the photometric redshift peaks of our candidate protoclusters (solid blue) as well as the distribution of the peaks themselves (dotted blue). In red is the redshift distribution of red IRAC galaxies in the lightcone. All distributions have been normalised by amplitude (i.e dividing by maximum bin count).}}
        \label{RedDistPeaks}
\end{figure}

The redshift distributions of the galaxies that fall within the peaks, as well as the redshifts of the peaks themselves, are shown in Figure~\ref{RedDistPeaks}. Here we can see that the distribution of these galaxies follows the distribution of galaxies selected by the IRAC colour cut fairly well, which explains why we have found peaks at $z<0.5$. We do, however, see a faster drop-off of high redshift ($z>1.5$) galaxies compared to the red IRAC distribution. We believe one reason for this is that galaxies at higher redshifts have larger photometric redshift errors (galaxies with $z_{phot} > 1.5$ have errors $\sim70\%$ larger than those with $z_{phot} < 1.5$). Larger errors hinder the search for photometric redshift overdensities, resulting in the identification of fewer redshift peaks. The distribution at lower redshifts ($z<1.3$) also does not match perfectly, and we appear to locate fewer protoclusters compared to the distribution of red IRAC galaxies. This can be attributed to the low completeness of $z<1.3$ galaxies as a result of the IRAC colour cut which makes the search for colour-selected galaxy overdensities more difficult, resulting in fewer detections.

\cite{Krefting2020} present 339 overdensities in the range $0.1 < z < 1.67$ in the XMMLSS field using photometric redshifts derived from \textit{u}-band through 4.5$\mu$m band photometry. As a quick comparison, we search for matches within a 2 arcminute radius. We find that 9 of our candidates (X4, X18, X19, X22, X43, X44, X49, X52 and X57) match with their overdensities ($\#125$, $\#322$, $\#319$, $\#315$, $\#321$, $\#250$, $\#320$, $\#280$ and $\#102$, respectively). The photometric redshifts of the matching overdensities, estimated by \cite{Krefting2020}, suggest that candidates X4 and X57 may be at low redshift ($z < 0.8$), while the rest are all at $z > 1$.

\begin{figure}   \includegraphics[width=\columnwidth]{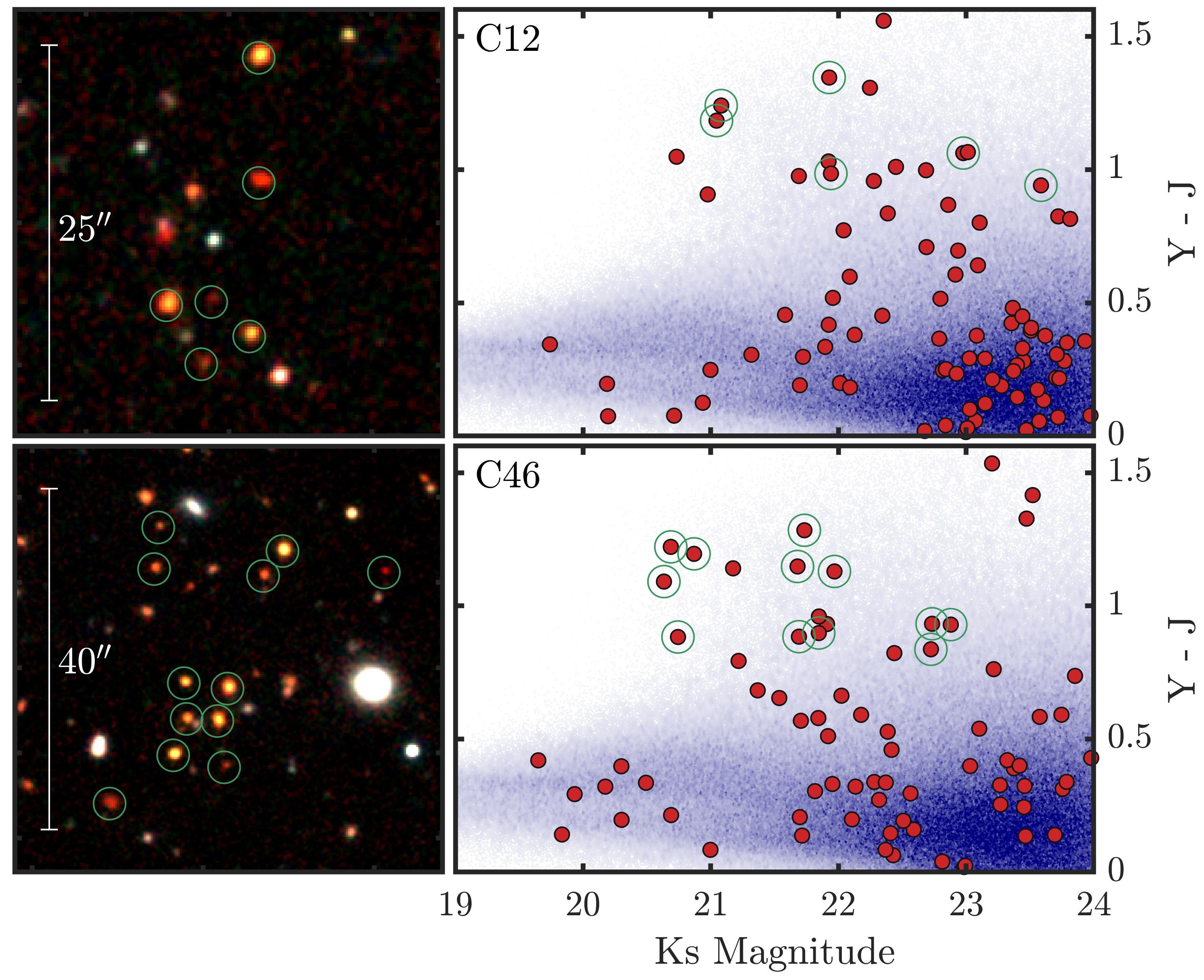}
    \caption{\small{\textit{Left}: $Y$, $J$, $Ks$ images of candidate protoclusters C12 and C46 from the VIDEO survey. The images only cover a fraction of the sizes of the candidate protoclusters ($\sim10\%$ and $\sim25\%$ respectively) to highlight the grouping of red galaxies from the corresponding CMD (green circles). \textit{Right}: Colour-magnitude diagrams where galaxies that lie within the projected boundaries of the candidate protoclusters are shown as red circles, with those in the corresponding composite images highlighted in green. The background colours show the density of objects across the colour-magnitude space for the whole VIDEO survey.}}
        \label{CMD}
\end{figure}

In Figure~\ref{CMD}, we show examples of false-colour composite images and colour-magnitude diagrams (CMDs) for two candidate protoclusters in our sample; C12 and C46. We use the $Y$, $J$ and $Ks$ bands for both the composite images and the CMDs, which come from the Visible and Infrared Survey Telescope for Astronomy (VISTA) Deep Extragalactic Observations (VIDEO) survey \citep{Jarvis2013}. The composite images show zoomed-in regions around groups of red galaxies (in $Y-J$), which are highlighted by green circles in both the CMDs and composite images. We can see that both structures have an overdensity of red galaxies in a relatively small region (less than 0.5 square arcminutes). We find photometric redshift peaks of $z_{phot} = 1.55, 1.74$ for C12, and $z_{phot} = 1.71$ for C46. The $Y-J$ colour of galaxies at these redshifts would span the 4000{\AA} break, so these colours may indicate old stellar populations that are typically associated with clusters.

\section{Discussion}
\label{Discussion}
\subsection{Comparison to X-ray selected clusters}
\label{XrayComparison}

It is possible to locate high redshift clusters using thermal Bremsstrahlung emission from the hot intracluster medium \citep[ICM; e.g.][]{Fassbender2011, Willis2018, Trudeau2020, Koulouridis2021}, but this selection technique is biased towards identifying only the most massive clusters due to the relationship between the X-ray luminosity and cluster mass. Here we compare our catalogue of protoclusters with two X-ray selected cluster catalogues that cover portions of the same fields used in this work. We use the first \citep{Koulouridis2021} to search for low-redshift X-ray detected contaminants and the second \citep{Trudeau2020} to locate high-redshift X-ray detected protoclusters.

Out of the 1,559 clusters presented in \cite{Koulouridis2021}, 36 fall in the same area covered by \citetalias{Lacy2021}, and they are all at $z < 1.1$. We search within a 2 arcminute radius and find 3 matches. These are candidates E11, X49 and X57 which match with XClass clusters 534, 20372 and 476 with separations 1.52$^{\prime}$, 1.90$^{\prime}$ and 1.32$^{\prime}$ respectively. These separations fall within the positional uncertainties of our detected structures, and so are likely true matches. They have spectroscopic redshifts of $z=0.221$, $z=0.055$ and $z=0.307$. We find a photometric redshift peak for X57 of $z_{phot}=0.28$ which matches with XClass 476 ($z=0.307$) fairly well. To determine the chance that this is a line-of-sight alignment, we measured the probability that three or more of our candidates would match to the sample of 36 cluster candidates from \citet{Koulouridis2021}. We find that there is only a $\sim9\%$ chance that these matches are random line-of-sight alignments. Hence, using this low-redshift cluster sample, we can rule out E11, X49 and X57 as protoclusters.

\cite{Trudeau2020} present 35 clusters in the XMMLSS field at $z_{phot}>0.8$ with 9 having $z_{phot}>1.3$. We find 3 matches within 2 arcminutes. These are candidates X52, X58 and X62 matching with clusters T-34 (JKCS 041), T-35 (3XLSS J022734.1-041021) and T-33 (3XLSS J022806.4-044803) with separations 0.21$^{\prime}$, 1.02$^{\prime}$ and 0.45$^{\prime}$ respectively. JKCS 041 (matched with X52) is a spectroscopically confirmed cluster with redshift $z=1.80$ \citep{Newman2014}, while the other two are `New candidate clusters' with redshift estimates of $z_{phot}=1.93$ (T-35) and $z_{phot}=1.79$ (T-33). There is a $\sim5\%$ chance that these matches are random line of sight alignments.

It is not easy to detect X-ray emission from protoclusters (especially those with low mass), which results in the small number of $z > 1.3$ candidates in the studies above. We therefore stack the X-ray signals from our clusters in order to search for a fainter signal, following the method of \cite{Willis2018}. The X-ray images we use come from the XMM-\textit{Spitzer} Extragalactic Representative Volume Survey (XMM-SERVS), which covers 3.2 deg$^2$ in ELAIS S1 and 4.6 deg$^2$ in CDFS \citep{Ni2021}, and 5.3 deg$^2$ in XMMLSS \citep{Chen2018}\footnote{Available at \url{https://personal.science.psu.edu/wnb3/xmmservs/xmmservs.html}}. We compute the soft band ([0.2-2] keV for ELAIS S1 and CDFS, and [0.5-2] keV for XMMLSS) count rate image for each field by subtracting the background map from the photon image and dividing by the exposure time. To reduce noise, we remove any pixel with an exposure time less than $25\%$ of the maximum exposure time in the given field. To further reduce noise, we also perform sigma-clipping, iteratively removing pixels more than $3\sigma$ from the mean count rate. All point sources are masked using circular masks with diameter $\sim2.5$ times the FWHM of the XMM-\textit{Newton} EPIC-pn \citep[European Photon Imaging Camera; ][]{Turner2001} point-spread-function of 15$^{\prime\prime}$ (i.e. a radius of 5 pixels from the point source). 

\begin{figure}   \includegraphics[width=\columnwidth]{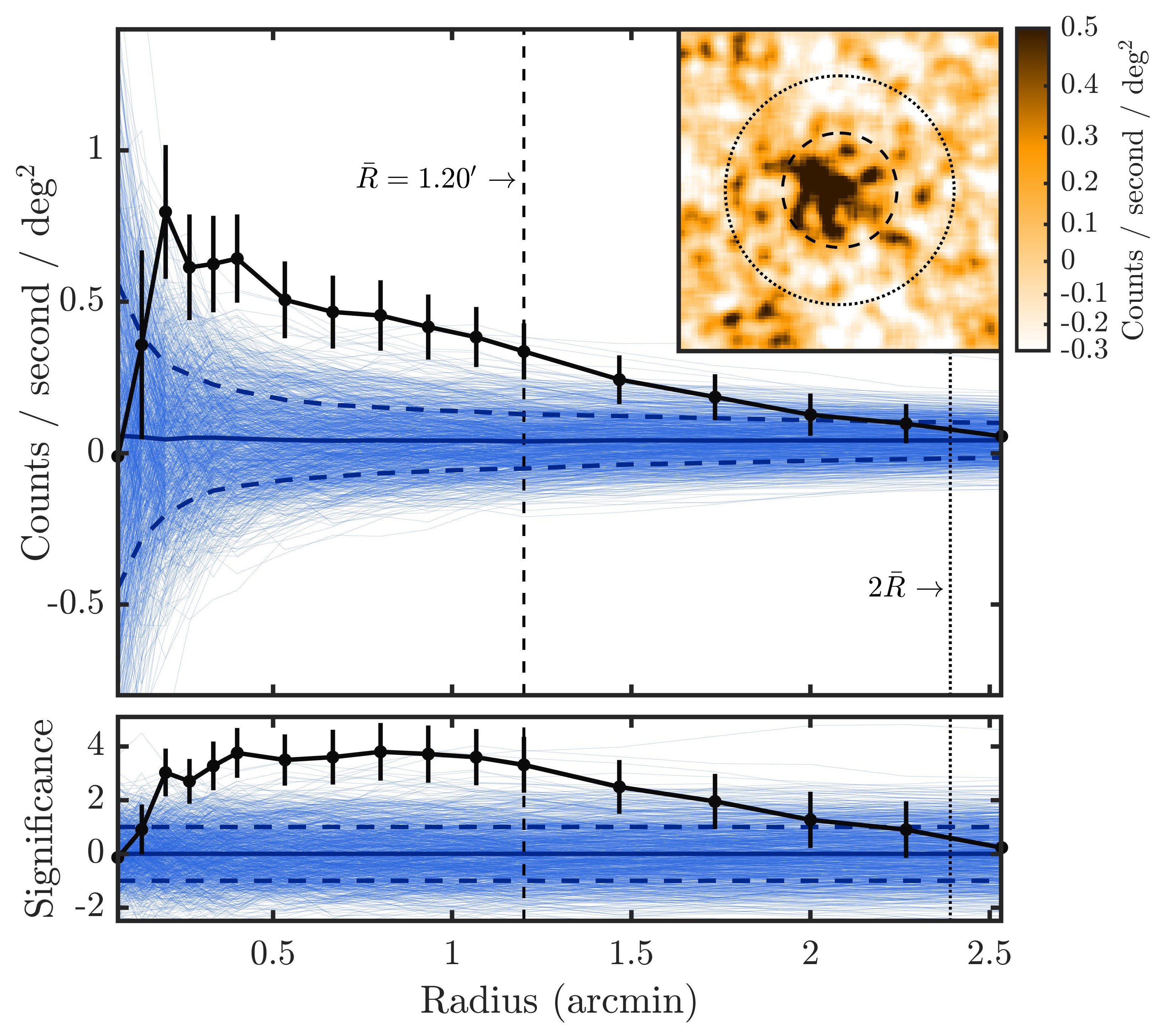}
    \caption{\small{\textit{Top}: The mean stacked X-ray signal from within a given radius for all of the candidate protoclusters that fall within the XMM-SERVS footprint (black line). The error bars on the black curve come from bootstrapping. There are 1,000 blue curves, each representing random stacked regions across the fields. The mean of the random stacks and one standard deviation from the mean are shown with the thicker blue solid and dashed lines respectively. \textit{Bottom}: The residuals from the top panels, representing the number of standard deviations from the mean. The vertical dashed black line represents the mean projected radii of the candidate protoclusters that have been stacked, assuming they are circular, with the dotted black line representing 2 times this value. \textit{Cutout}: The stacked X-ray image, smoothed with a Gaussian kernal with standard deviation of width $\sim9^{\prime\prime}$ and exponentially scaled. The dotted and dashed circles represent the same radii as in the main plot.}}
        \label{XrayMapStack}
\end{figure}

Square regions of $101\times101$ pixels are centered on our protocluster positions (from Table~\ref{TabGalsE}), only keeping those that fall within the XXM-SERVS footprint entirely. Out of the 189 regions, 146 fall within the XXM-SERVS footprint (with 118 actually having some X-ray signal). Each of these regions are stacked on top of one another, with the mean count rate along each pixel calculated (excluding NaNs). The final smoothed stacked X-ray image is shown in the inset plot of Figure~\ref{XrayMapStack}, where we have used exponential scaling to highlight the signal. In the main part of same figure we also present the unsmoothed X-ray radial profile in black. To determine the robustness of this signal, we perform a bootstrap analysis which involved randomly resampling each of the regions 1,000 times, allowing for the repeated selection of individual regions. From this, we have calculated the error bars shown in Figure~\ref{XrayMapStack}. We also stack random regions within each field (equal in number to the protocluster regions in each field), in order to determine the significance of our signal. The blue lines in Figure~\ref{XrayMapStack} show the radial profile of X-rays for each iteration of the stacking of random regions. We do this 1,000 times, with the mean and standard deviation of the radial profiles also shown. Computing the significance ($(S-\mu) / \sigma$), we can see that the stacked X-ray signal within the mean effective radius of our protocluster sample ($1.2^{\prime}$) is almost at a significance of $4\sigma$, with the bootstrapping analysis suggesting that a significant number of our candidates are in collapsed halos.

\begin{figure}   \includegraphics[width=\columnwidth]{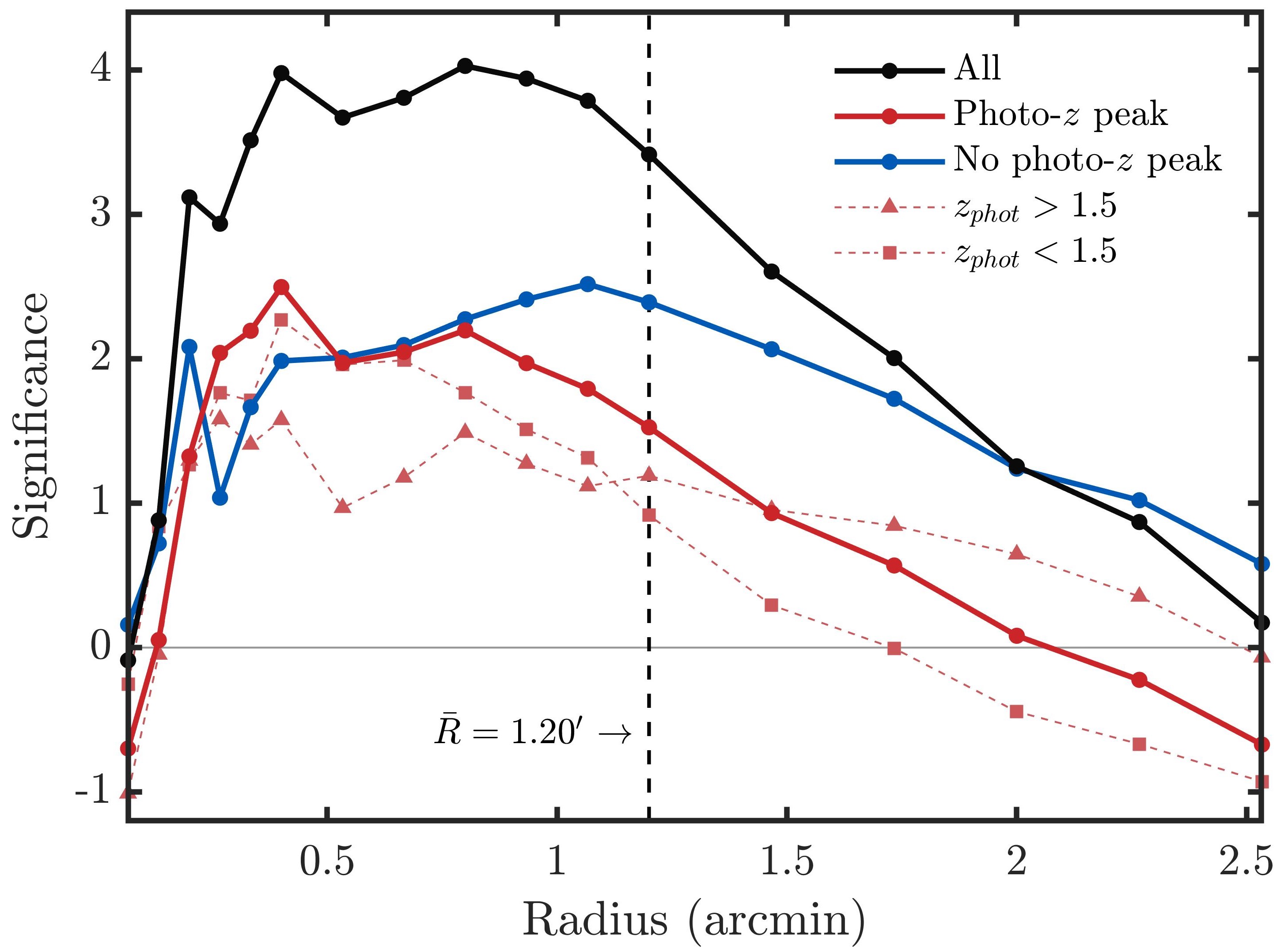}
    \caption{\small{The X-ray stacking residuals of the different subsamples of our candidate protoclusters within a given annulus (akin to the bottom panel of Figure~\ref{XrayMapStack}). The solid curves represent all candidate protoclusters (black), those with a photometric redshift peak (red) and those without (blue). The dashed red lines correspond to high (triangles) and low (square) redshift subsamples of those with photometric redshift peaks.}}
        \label{XrayResid}
\end{figure}

With our estimates of the photometric redshift of each protocluster, we can stack different redshift subsamples. We therefore perform the method outlined above on all candidate protoclusters that we have detected a redshift peak for (further split into $z_{phot} > 1.5$ and $z_{phot} < 1.5$) as well as those without (only if they fall within footprint of photometric redshift catalogues). For those that have multiple peaks with at least one above $z_{phot} > 1.5$ and one below, we include in both subsamples. The significance of each are shown in Figure~\ref{XrayResid}. Comparing those with photometric redshift peaks and those without, we can see the signal is largely the same within $1.2^{\prime}$, but significantly different at larger radii. One explanation of this could be that it is harder to detect a photometric redshift peak at higher redshifts due to the larger uncertainties, meaning the majority of protoclusters that make up this subsample are potentially at $z>2$, and therefore are less likely to have collapsed. In such a system, there may be an X-ray signal from multiple nonconcentric halos extending the X-ray signal to higher radii. It must be noted that the significance of these signals and their differences are fairly low, and so are by no means conclusive. However, the fact that we still have a $2\sigma$ X-ray detection for those that we could not find a photometric redshift peak for, suggests there may in fact be clusters there, and that photometric redshift overdensity searches are not complete. If we now compare the high and low redshift signals, we find that the X-ray signal within $1.2^{\prime}$ is dominated by $z_{phot} < 1.5$ protoclusters, whereas $z_{phot} > 1.5$ protoclusters dominate at higher radii. This could again be explained by the fact that protoclusters at higher redshifts are are made of multiple nonconcentric haloes, each emitting X-rays at a significant distance from what we define as the cluster centre.

\subsection{Comparison to spectroscopically confirmed high-redshift clusters}
\label{ClusterComparison}

In addition to JKCS 041 at $z=1.8$ (labelled X52 in our catalogue), there are a few other well-known structures in the literature that are within the DDFs above a redshift of 1.3. The initial detection of these structures were through varying methods, such as within the \textit{Spitzer} Adaptation of the Red Sequence Cluster survey \citep[SpARCS;][]{Muzzin2009, Wilson2009}, but they have all since been spectroscopically confirmed. Their properties are shown in Table~\ref{SpecClusters}.

\begin{figure*}
\includegraphics[width=\textwidth]{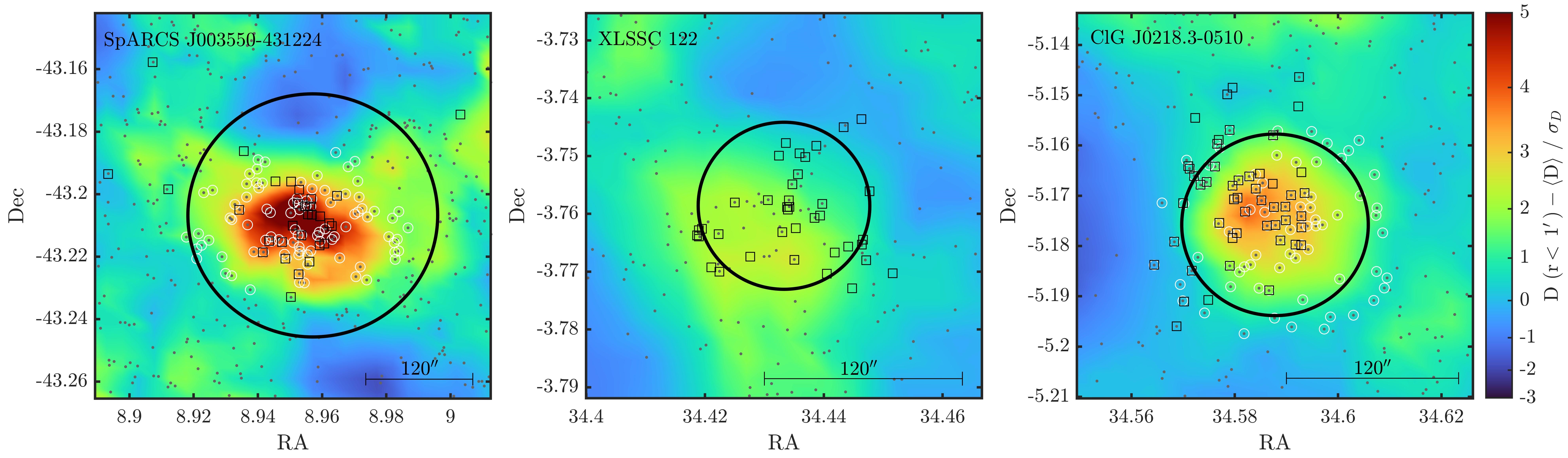}
    \caption{\small{Spectroscopically confirmed protoclusters in ELAIS S1 (\textit{left}) and XMMLSS (\textit{middle} and \textit{right}). Black squares represent spectroscopically confirmed members with the black circles representing $r_{200}=1.2$ Mpc ($140^{\prime\prime}$ at $z=1.34$), $r_{200}=443$ kpc ($52^{\prime\prime}$ at $z=1.98$), and $r_{200}=560$ kpc ($65^{\prime\prime}$ at $z=1.62$), respectively. Grey dots are the red IRAC galaxies in the fields with those that have been selected by $D (r<1^{\prime}) - \langle D \rangle / \sigma_D > 4.25$ displayed as white cirlces. The background map is the smoothed density map of red IRAC galaxies.}}
        \label{Map}
\end{figure*}

We find five matches to the structures in our catalogue within 2 arcminutes, which are shown in Table~\ref{SpecClusters}. We show two examples of structures we do detect (SpARCS J0035-4312, ClG J0218.3-0510) and one we do not (XLSSC 122) in Figure~\ref{Map}. This figure shows the spectroscopically confirmed members of these structures, their respective radii, and the surrounding red IRAC galaxies. Also highlighted in white circles are the red IRAC galaxies that have been selected via our method (which belong to groups E20 and X9, respectively), demonstrating the method's feasibility. In fact, we even recover the confirmed spectroscopic redshift of SpARCS J0035-4312 in our photometric redshift analysis.

There are a number of structures we do not detect, but this is a result of our inclination towards higher purity values, at the expense of completeness. We found that it is possible to detect some of the structures we miss if we use different parameter values. For example, we can detect XLSSC 122 if we use a search radius of $0.5^{\prime}$, but this would increase the contamination of the overall sample. We therefore compromise our completeness in order to produce as pure a sample as possible.

\begin{table*}
\centering
\caption{Spectroscopically confirmed clusters and protoclusters above $z > 1.3$ within the DDFs, ordered by redshift. We also list the protoclusters detected in this work (from Table~\ref{TabGalsE}) whose positions match with these structures within 2 arcminutes.}
\begin{tabular}{wc{3cm} wc{1.2cm} wc{1.2cm} wc{1.2cm} wc{1.2cm} wc{1.2cm} >{\centering\arraybackslash}p{5.2cm}c}
\hline
Name & Matched with & RA & Dec & Redshift & M$_{200}$ & Sources \\
 & (separation) & & & & ($10^{14}$ M$_{\odot}$) & \\
\hline \\ [-2ex]
SpARCS J0219-0531 & - & 34.9315 & -5.5249 & 1.325 & 2.51$^{+\,1.33}_{-0.98}$  & \cite{Wilson2009,Chan2021} \\
SpARCS J0035-4312 & E20 ($15^{\prime\prime}$) & 8.9570 & -43.2066 & 1.34 & $9.4\pm6.2$ & \cite{Wilson2009,Balogh2021}\\
SpARCS J0335-2929 & - & 52.7649 & -29.4821 & 1.369 & 1.60$^{+\,0.65}_{-0.51}$ & \cite{Nantais2016,Chan2021} \\
SXDF87XGG & - & 34.5360 & -5.0630 & 1.406 & $0.77\pm0.10$ & \cite{Finoguenov2010,Balogh2021} \\
SXDF76XGG & - & 34.7461 & -5.3041 & 1.459 & $0.86\pm0.19$ & \cite{Finoguenov2010,Balogh2021} \\
SpARCS J0225-0355 & X47 ($45^{\prime\prime}$) & 36.4399 & -3.9214 & 1.598 & - & \cite{Wilson2009,Nantais2016} \\
ClG J0218.3-0510 & X9 ($55^{\prime\prime}$) & 34.5750 & -5.1667 & 1.62 & $0.77\pm0.38$ & \cite{Papovich2010,Papovich2012,Tanaka2010,Pierre2012}\\
SpARCS J0330-2843 & - & 52.7330 & -28.7165 & 1.626 & 2.4$^{+\,1.0}_{-1.5}$ & \cite{Lidman2012,Muzzin2013b}\\[0.3ex]
SpARCS J0224-0323 & X41 ($70^{\prime\prime}$) & 36.1097 & -3.3919 & 1.633 & 0.4$^{+\,0.1}_{-0.3}$ & \cite{Lidman2012,Muzzin2013b}\\
JKCS 041 & X52 ($29^{\prime\prime}$) & 36.6817 & -4.6893 & 1.803 & $1.8\pm1.7$ &  \cite{Andreon2009,Newman2014,Andreon2014}\\
XLSSC 122 & - & 34.4333 & -3.7586 & 1.98 & $1.9\pm2.1$ & \cite{Willis2013,Willis2020,Mantz2018} \\
\hline \\
\end{tabular}
\label{SpecClusters}
\end{table*}

\subsection{Biases}
\label{Biases}

From the comparison to other protocluster and cluster catalogues in the section above, we see that the IRAC-selected protocluster candidates are a biased subsample of all the (proto)clusters in the field. To understand how our selection criteria bias the protocluster sample we perform our detection method on the lightcone and compare properties of the protoclusters that we select to those we do not. As there is a level of randomness involved with preparing the lightcone for the detection method (see section \ref{Simulations}), we run the method 500 hundred times. 

\begin{figure}   \includegraphics[width=\columnwidth]{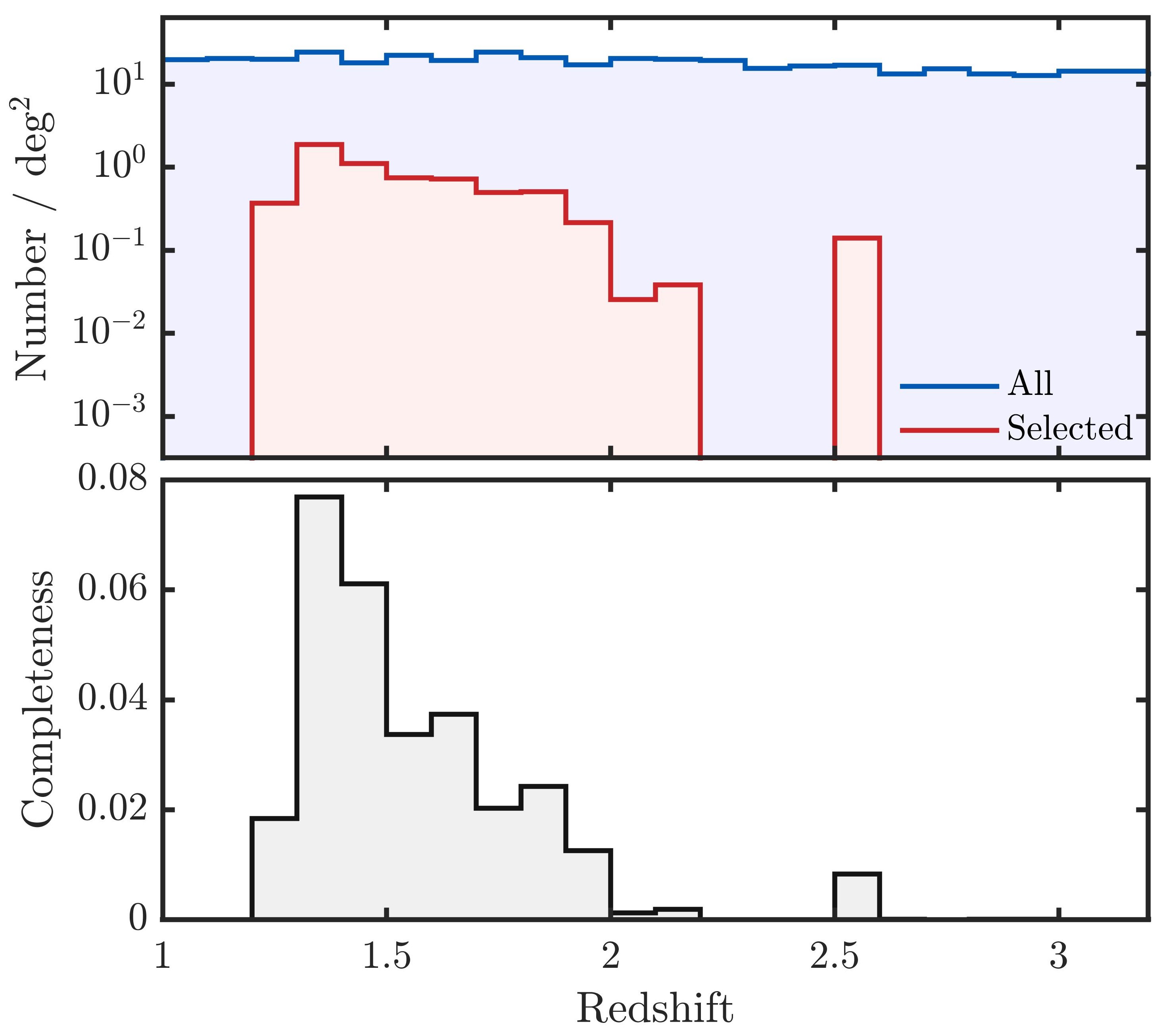}
    \caption{\small{\textit{Top}: The redshift distribution of all protoclusters in the lightcone (blue) and the mean redshift distribution of those that are selected by our method over 500 iterations (red). \textit{Bottom}: The completeness of selected protoclusters as a function of redshift.}}
        \label{RedComp}
\end{figure}

Out of the 1,789 $1<z<5$ protoclusters within the lightcone (of which 1,070 are within $1.3 < z < 3.2$) , we select (on average) just 19 of them using our optimal selection criteria. Figure~\ref{RedComp} shows the redshift distribution of protoclusters in the lightcone, as well as the average redshift distribution of the protoclusters we select with the IRAC method (only the successful detections). In the bottom panel of the same figure is the completeness as a function of redshift. We can see that the vast majority of protoclusters we select are in the redshift range $1.2<z<2$, with a very small minority at higher redshifts. We therefore limit our bias analysis to this redshift range $1.2<z<2$. 

We checked whether the magnitude limit we use affects the redshifts of the selected structures by reproducing Figure~\ref{RedComp} for incremental depths up to 25 mag. We found that no matter how deep the data (up to 25 mag), we were still limited to structures within $z\lesssim2$. This is likely due to the fact that the deeper data results in many more faint $z < 2$ galaxies, which increases the overdensity threshold, meaning the $z > 2$ structures do not have densities that are significant enough to be identified. Therefore the \textit{Spitzer}/IRAC method for selecting protolcusters is only efficient up to $z=2$, even though in principle the [3.6] - [4.5] colour cut can select galaxies up to $z=3.2$.

\begin{figure}   \includegraphics[width=\columnwidth]{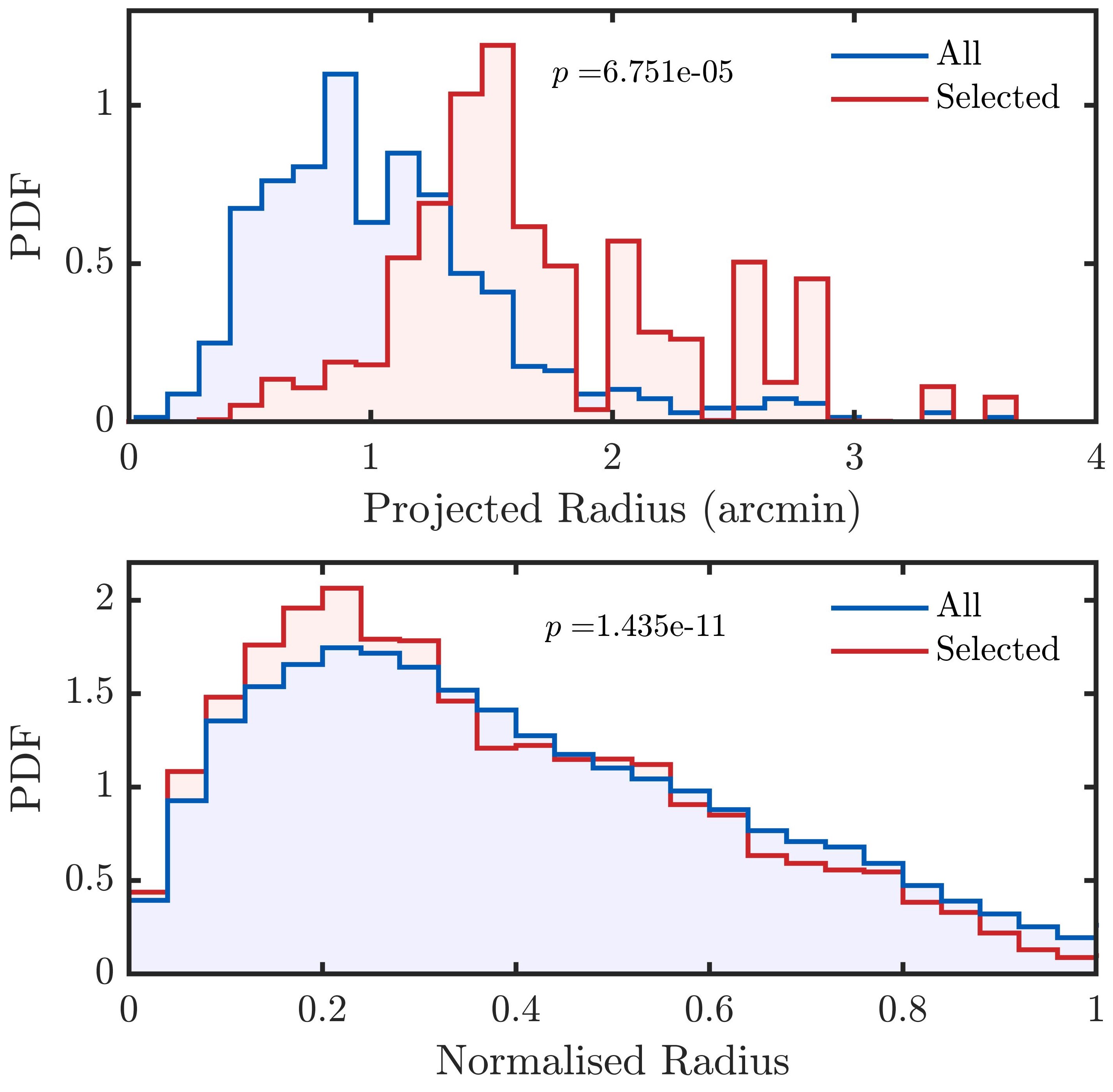}
    \caption{\small{\textit{Top}: The distribution of projected radii of all $1.2 < z < 2$ protoclusters in the lightcone (blue) and the mean projected radii distribution of those that are selected by our method over 500 iterations (red). \textit{Bottom}: The radial distribution of galaxies in all $1.2 < z < 2$ protoclusters in the lightcone (blue), normalised to their maximum radius, and the mean radial distribution of galaxies in those that are selected by our method over 500 iterations (red).}}
        \label{SizeCompact}
\end{figure}

To understand the biases of our sample, we compare properties of the protoclusters, including size, compactness, richness, halo mass, and its descendant $z=0$ halo mass. We define the projected radius of a protocluster as the circularised radius, from the area within the projected conforming boundary of the member galaxies. The distributions of these sizes are shown in the top panel of Figure~\ref{SizeCompact} which show that the IRAC method tends to select protoclusters that are larger in size than the general population. This is confirmed quantitatively via the two-sample KS test, where a p-value of $6.751\times10^{-5}$ is obtained.

While the optimised IRAC method tends to select larger structures, it also tends to select structures that are more centrally concentrated. This is shown in the bottom panel of Figure~\ref{SizeCompact}, where we plot the radial distributions of galaxies in protoclusters, normalised to their maximum radius. Here we can see that the method selects protoclusters whose galaxies are skewed more towards their centres. The KS test returns a p-value of $1.435\times10^{-11}$, again showing the significance of the difference between the two distributions.

\begin{figure*}   \includegraphics[width=\textwidth]{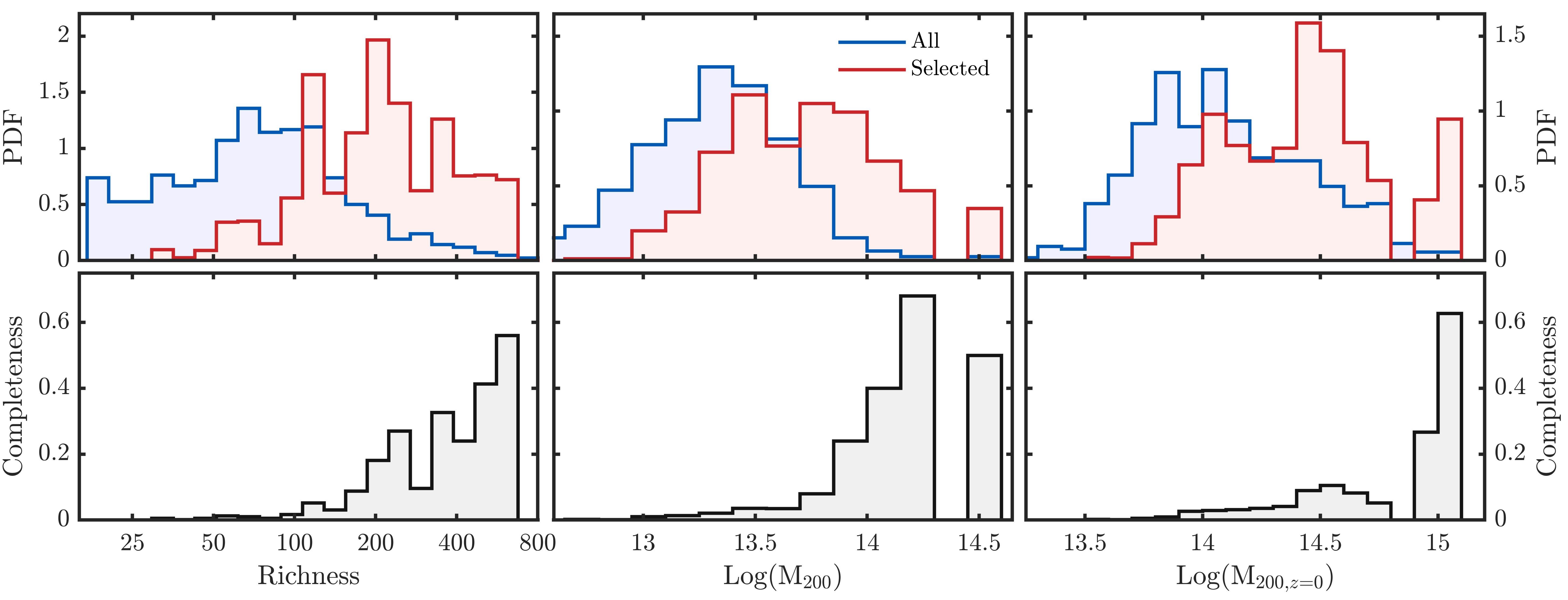}
    \caption{\small{\textit{Left}: The $z=0$ halo mass distribution for all $1.2 < z < 2$ protoclusters in the lightcone (blue) and the mean halo mass distribution of those that are selected by our method over 500 iterations (red). We also show the completeness as a function of $z=0$ halo mass below. \textit{Middle}: The halo mass distribution for all $1.2 < z < 2$ protoclusters in the lightcone (blue) and the mean halo mass distribution of those that are selected by our method over 500 iterations (red). We also show the completeness as a function of halo mass below. \textit{Right}: The richness distribution for all $1.2 < z < 2$ protoclusters in the lightcone (blue) and the mean richness distribution of those that are selected by our method over 500 iterations (red). We also show the completeness as a function of richness below.}}
        \label{HMRich}
\end{figure*}

In the upper left panel of Figure~\ref{HMRich}, we plot the richness distributions. The distributions of those that we select versus those we do not are almost the inverse of one another, showing how the IRAC method is biased to select the richest protoclusters. In fact, if we look at the completeness as a function of richness in the bottom left panel, we see that we only detect a tiny number (less than $1\%$) of structures with fewer than 100 members. However, for clusters with more than 500 member galaxies, the optmised method is $40\%$ complete (over 10 times higher than the total completeness for $1.2<z<2$ protoclusters). 

The most massive halo in the selected protoclusters is more massive than for the general population of protoclusters. This is shown in the middle panel of Figure~\ref{HMRich} where we plot M$_{200}$ of the most massive halo within the selected protocluster. Almost all of the selected protoclusters already contain a group or cluster-mass halo. While group-mass halos are also common in the whole protocluster population, they are generally 0.5 dex less massive than in the selected protoclusters.

We finally compare the $z=0$ halo masses of the protoclusters we select. The way we have defined protoclusters (section~\ref{Optimisation}) allows galaxies from the same protoclusters to end up in different $z=0$ halos. Therefore, we take the weighted average of the $z=0$ halo mass of each galaxy in a protocluster to give the final $z=0$ halo mass for that protocluster. The distributions for these halo masses are shown in the upper right panel of Figure~\ref{HMRich}, where we show that the IRAC method tends to select protoclusters that form higher mass halos by $z=0$. The panel below shows the completeness as a function of halo mass -- showing the method is $50\%$ complete for M$_{200,z=0} > 10^{14.9}$ M$_{\odot}$.

Overall, we find that protoclusters selected by the \textit{Spitzer}/IRAC method are heavily biased towards larger, richer, more massive, and more centrally concentrated protoclusters, that will evolve into more massive clusters by $z=0$. This inclination towards specific properties may result in a bias in the observed properties, such as quenched fractions, and other galaxy scaling relations measured from this biased protocluster sample. This may then affect the number and type of supernovae observed from this sample, hence any interpretation of this sample must take into account the cluster sample biases. 

\section{Conclusions}
\label{Conclusions}

We identify 189 candidate protoclusters in three of LSST's deep drilling fields, covering an area of around 30 square degrees. This sample was selected using a \textit{Spitzer} IRAC red colour-cut to identify $z>1.3$ galaxy overdensities. The selection criteria were chosen by optimising the purity of the selected protocluster sample, as measured on a lightcone that was matched to the IRAC data available on the deep drilling fields. Based on the lightcone testing, we estimate that between $60\%$ and $80\%$ of the candidates are likely genuine protoclusters. This assertion is corroborated by a robust $\sim4\sigma$ stacked X-ray signal originating from these structures. We bolstered the information we have on these structures by searching for photometric redshift peaks, where for 47 of them we found a redshift peak at $z_{phot} > 1.1$.

The purpose of this study was to identify regions of the deep drilling fields which are likely to have supernovae that are hosted by protocluster members. The positional uncertainty of our protocluster catalogue is $\sim2$\,arcmin (from the lightcone tests), and the typical size of the protoclusters is 1.5\,arcmin. We therefore suggest that transient sources in the $z$ or $y$ bands (which are not visible in the bluer optical bands), and are within 3.5\,arcmin of the 189 candidates, are potential supernovae of protocluster members that are likely to be at $1<z<2$. Future measurement of the supernovae rate and supernovae types can illuminate the star formation and metal enrichment history of clusters during their early assembly period. 

Identifying protoclusters as overdensities of \textit{Spitzer}/IRAC colour-selected galaxies has been one of the most widely employed protocluster detection method and we have used the lightcone to explore the purity of various protocluster samples in the literature. We find that samples selected from shallow observations ([4.5] $<$ 22 mag) or at relatively low overdensity significance (e.g. $\sim$2 $\sigma$) resulted in highly contaminated samples of protocluster candidates. These samples had purities of 30-40\%. We furthermore show that including an optical magnitude cut (e.g. $I < 20.45$ mag) does not improve the sample purity, but taking a $z^{\prime}$ - $3.6\mu$m colour cut once the LSST data is available will increase the sample purity to $\sim82\pm17\%$. The optimal parameters for identifying a highly pure sample of protoclusters using \textit{Spitzer} IRAC data is  using data of at least [4.5]$\sim$22 mag depth (but more depth does not produce purer or higher redshift samples), overdensities of at least 4$\sigma$ significance measured in apertures of $1$\,arcmin radius and with galaxies redder than [3.6] -[4.5] $> -0.05$ (although the range -0.2 to 0 also works just as well). 

We also show that \textit{Spitzer}-selected overdensities are only able to efficiently select protoclusters at $1<z<2$. Even though the method works, in principle, out to $z=3.2$, the overdensities at $z>2$ tend to be of too low significance to be selected whilst also ensuring the sample has reasonable level of purity. We therefore recommend that alternative protocluster detection methods should be employed to locate protoclusters at $z>2$ in the deep drilling fields, such as searching for overdensities of Lyman-break galaxies. 

To obtain the purest possible sample, the method produces a highly incomplete sample -- accounting for only $\sim4\%$ of the actual population of protoclusters. Furthermore, the sample exhibits a pronounced bias towards larger, more massive, and centrally concentrated protoclusters that form more massive clusters at $z=0$. Hence any future study of this, or other \textit{Spitzer}-selected protocluster samples, must note that the protocluster members may be biased relative to the whole population of protocluster members due to this selection bias.

\section*{Acknowledgements}
This work was developed as part of an ISSI workshop on protoclusters, held in Bern, Switzerland in 2022. We are grateful for the support of ISSI and the use of their facilities. We also gratefully acknowledge the Lorentz Center in Leiden (NL) for facilitating discussions on this project. HG acknowledges support through an STFC studentship. NAH acknowledges support from STFC through grant ST/X000982/1. BV acknowledges support from the INAF Mini Grant 2022 "Tracing filaments through cosmic time" (PI Vulcani). GW gratefully acknowledges support from the National Science Foundation through grant AST-2205189 and from HST program number GO-16300. Support for program number GO-16300 was provided by NASA through a grant from the Space Telescope Science Institute, which is operated by the Association of Universities for Research in Astronomy, Incorporated, under NASA contract NAS5-26555. We are also grateful to the reviewer for their prompt and insightful feedback.

For the purpose of open access, the author has applied a creative commons attribution (CC BY) to any author accepted manuscript version arising.

\section*{Data Availability}

The observational data used in this paper are publicly available from the following locations:

\noindent \textit{Spitzer}/IRAC -- \url{https://irsa.ipac.caltech.edu/data/SPITZER/DeepDrill/overview.html} 

\noindent X-ray images -- \url{https://personal.science.psu.edu/wnb3/xmmservs/xmmservs.html} 

\noindent CDFS and ELAIS S1 photo-$z$ catalogue -- \url{https://zenodo.org/records/4603178} 

\noindent XMMLSS photo-$z$ catalogue -- \url{https://vizier.cds.unistra.fr/viz-bin/VizieR?-source=J/MNRAS/513/3719} 

\noindent VIDEO Survey -- \url{http://eso.org/rm/publicAccess#/dataReleases}.

The lightcone is available by request from Micol Bolzonella. Scripts to produce the results presented here are available by request from the corresponding author. 


\bibliographystyle{mnras}
\bibliography{refs} 




\appendix

\section{Identification of cluster progenitors}
\label{Appendix}

Galaxy clusters in the lightcone are identified exclusively on dark matter halo mass. Any Friends-of-Friends (FoF) halo with M$_{200}$ / M$_{\odot}\,\geq 10^{14}$ at $z=0$ is defined as a cluster. The merger trees of these halos can be traced back to any redshift in order to identify their progenitors. All galaxies associated with these progenitor halos are identified as cluster progenitor members. Using this definition, we find 789,509 cluster progenitor members contained within 3,908 unique cluster progenitors.

In this set of 3,908 cluster progenitors, we find that some have unrealistic properties; specifically, some have unrealistic extents while others have very few members. These unrealisitc properties can arise as an artifact of the lightcone creation, where the the simulation box has been cut - meaning some fraction of the member galaxies of a cluster progenitor end up placed in a different part of the lightcone or where structures get cut leaving only a handful of members from a particular cluster progenitor.

For each cluster progenitor in the lightcone, a maximum redshift extent is calculated using the highest and lowest redshifts of member galaxies. We find that 140 out of 3,908 cluster progenitors have unrealistic redshift extents of more than 1.5. We split split these cluster progenitors into two and refer to each as a unique cluster progenitor. This leaves us with a set of 4,048 unique cluster progenitors. 

\begin{figure}
    \includegraphics[width=\columnwidth]{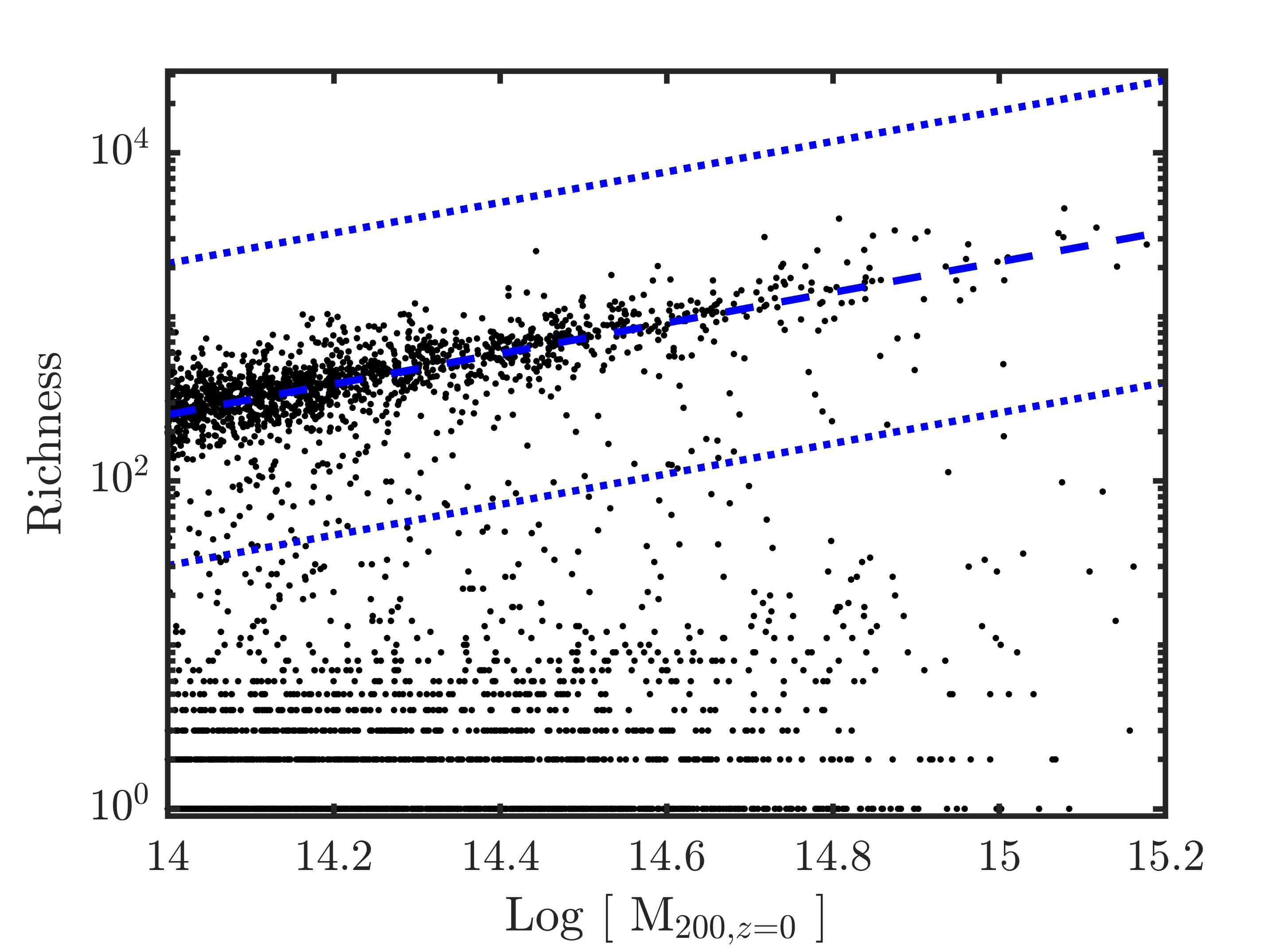}
    \caption{\small{The mass-richness relation of the set of 4,048 unique protoclusters (black points) with the robust fit (dashed blue line) and $5\sigma$ error on the robust fit (dotted blue lines).}}
        \label{MRR}
\end{figure}

From the resulting set of cluster progenitors, we can find a mass-richness relation in order to identify any remaining problematic cases. In Figure~\ref{MRR}, we can see there are a significant number of cluster progenitors with unrealistically few members (e.g. $N\,<\,5$). There is a clear relationship between the $z=0$ halo mass and the number of cluster progenitor members. Therefore, we use an iteratively reweighted least squares method to fit a linear regression model, on those cluster progenitors with 5 or more members. Initially, each data point is assigned equal weight, and the algorithm estimates the model coefficients using ordinary least squares. After each iteration, the algorithm computes the weights of each data point, giving lower weight to points farther from model predictions in the previous iteration until the values of the coefficient estimates converge within a specified tolerance. We find that 2,479 unique cluster progenitors containing 10,042 galaxies are more than $5\sigma$ away from the robust fit. We remove these cluster progenitors and their members from our list, leaving us with 779,467 galaxies within 1,569 cluster progenitors. This removal of $61\%$ of cluster progenitors only corresponds to $1.3\%$ of cluster progenitor galaxies. Any subsequent mention of cluster progenitors within the lightcone will be referring to this list of 779,467 galaxies within 1,569 cluster progenitors only, all other galaxies previously referred to are now considered field galaxies.

\begin{table*}
\centering
\caption{Candidate protoclusters in the CDFS, ELAIS S1, and XMMLSS Deep Drilling Fields. Near-infrared data is vital to identify and classify supernovae at $z>1$, so we highlight the candidates that also fall into the expected observing area of the Euclid deep fields and auxiliary calibration fields with an asterisk (apart from candidates in the CDFS as they all fall within these fields).}
\begin{tabular}{wc{3cm} wc{1.2cm}  wc{1.2cm} wc{2cm} wc{2cm} wc{3cm}}
\hline
Group ID$^\textrm{\textit{a}}$ & RA$^\textrm{\textit{b}}$ & Dec$^\textrm{\textit{b}}$ & Size$^\textrm{\textit{c}}$ & Best $z_{phot}$ & Number Galaxies in \\
(AKA) & & & (arcmin$^2$) & estimate$^\textrm{\textit{d}}$ & Redshift Range$^\textrm{\textit{d}}$ \\
\hline \\ [-2ex]
C1 & 51.4059 & -29.0585 & 4.28 & N/A & N/A \\
C2 & 51.5579 & -27.6287 & 6.16 & N/A & N/A \\
C3 & 51.5773 & -28.0976 & 2.72 & N/A & N/A \\
C4 & 51.7520 & -27.2838 & 3.96 & - & - \\
C5 & 51.7800 & -27.3562 & 7.80 & 1.46$^{+\,0.04}_{-0.06}$, 1.58$^{+\,0.02}_{-0.03}$, 1.76$^{+\,0.09}_{-0.11}$ &  40, 12, 43 \\
C6 & 51.7902 & -28.6552 & 2.41 & 1.68$^{+\,0.02}_{-0.03}$, 1.84$^{+\,0.11}_{-0.09}$ &  7, 17 \\
C7 & 51.8621 & -28.5197 & 2.44 & 1.45$^{+\,0.10}_{-0.10}$ & 23 \\
C8 & 51.8665 & -29.1094 & 4.72 & - & - \\
C9 & 51.9194 & -27.6045 & 3.97 & 1.37$^{+\,0.08}_{-0.07}$, 1.61$^{+\,0.04}_{-0.06}$, 2.03$^{+\,0.02}_{-0.03}$ &  20, 8, 4 \\
C10 & 51.9970 & -27.5911 & 6.74 & 1.43$^{+\,0.02}_{-0.03}$ & 22 \\
C11 & 52.1155 & -28.0637 & 2.60 & 1.61$^{+\,0.09}_{-0.11}$ & 18 \\
C12 & 52.2003 & -28.1125 & 2.69 & 1.55$^{+\,0.05}_{-0.05}$, 1.74$^{+\,0.11}_{-0.09}$ &  9, 20 \\
C13 & 52.2078 & -27.7793 & 2.62 & 1.50$^{+\,0.05}_{-0.05}$ & 14 \\
C14 & 52.2384 & -26.8898 & 2.49 & N/A & N/A \\
C15 & 52.2531 & -27.1176 & 4.91 & 1.88$^{+\,0.02}_{-0.03}$ & 4 \\
C16 & 52.2779 & -27.5723 & 3.32 & 1.58$^{+\,0.07}_{-0.08}$, 1.80$^{+\,0.10}_{-0.10}$ &  11, 13 \\
C17 & 52.3316 & -28.4975 & 7.34 & 2.04$^{+\,0.11}_{-0.09}$ & 21 \\
C18 & 52.3949 & -29.5898 & 2.57 & N/A & N/A \\
C19 & 52.4083 & -27.6060 & 2.55 & 0.03$^{+\,0.02}_{-0.03}$, 1.43$^{+\,0.02}_{-0.03}$, 1.59$^{+\,0.11}_{-0.09}$ &  6, 6, 19 \\
C20 & 52.4144 & -27.0010 & 2.21 & - & - \\
C21 & 52.4287 & -29.6724 & 5.30 & N/A & N/A \\
C22 & 52.5279 & -27.7424 & 4.92 & 1.83$^{+\,0.02}_{-0.03}$, 1.93$^{+\,0.02}_{-0.03}$, 2.03$^{+\,0.02}_{-0.03}$ &  4, 8, 4 \\
C23 & 52.7175 & -28.9302 & 8.51 & 1.93$^{+\,0.02}_{-0.03}$ & 7 \\
C24 & 52.7288 & -28.7900 & 5.02 & - & - \\
C25 & 52.7838 & -28.7139 & 16.10 & 1.59$^{+\,0.06}_{-0.04}$ & 29 \\
C26 & 52.7846 & -27.3995 & 6.13 & 1.28$^{+\,0.02}_{-0.03}$, 1.56$^{+\,0.14}_{-0.11}$ &  10, 70 \\
C27 & 52.8079 & -28.0439 & 2.63 & 1.39$^{+\,0.06}_{-0.04}$, 1.53$^{+\,0.02}_{-0.03}$ &  12, 8 \\
C28 & 52.8385 & -26.6153 & 6.64 & N/A & N/A \\
C29 & 52.8404 & -26.8564 & 2.31 & N/A & N/A \\
C30 & 52.8594 & -28.7577 & 2.44 & 1.58$^{+\,0.02}_{-0.03}$, 1.79$^{+\,0.06}_{-0.09}$, 1.95$^{+\,0.05}_{-0.05}$ &  4, 8, 15 \\
C31 & 52.9451 & -28.8012 & 4.16 & 2.00$^{+\,0.10}_{-0.15}$ & 15 \\
C32 & 53.0694 & -29.3069 & 2.40 & N/A & N/A \\
C33 & 53.0941 & -26.8814 & 4.71 & N/A & N/A \\
C34 & 53.1733 & -26.8156 & 2.69 & N/A & N/A \\
C35 & 53.2570 & -26.8720 & 2.85 & N/A & N/A \\
C36 & 53.3626 & -27.0511 & 5.36 & - & - \\
C37 & 53.3971 & -29.3774 & 2.83 & N/A & N/A \\
C38 & 53.4142 & -29.0578 & 5.10 & N/A & N/A \\
C39 & 53.5017 & -27.6560 & 5.03 & - & - \\
C40 & 53.6771 & -29.0376 & 2.64 & N/A & N/A \\
C41 & 53.6905 & -28.0530 & 6.24 & 1.40$^{+\,0.05}_{-0.05}$ & 27 \\
C42 & 53.6968 & -28.4207 & 2.49 & 1.93$^{+\,0.02}_{-0.03}$ & 5 \\
C43 & 53.7425 & -28.9336 & 11.15 & N/A & N/A \\
C44 & 53.7426 & -29.3748 & 2.93 & N/A & N/A \\
C45 & 53.8124 & -28.6257 & 3.77 & - & - \\
C46 & 53.8464 & -27.9388 & 2.70 & 1.71$^{+\,0.19}_{-0.21}$ & 31 \\
C47 & 53.8973 & -28.8443 & 5.14 & N/A & N/A \\
C48 & 53.9091 & -28.5428 & 13.28 & - & - \\
C49 & 53.9245 & -28.2513 & 2.87 & 1.93$^{+\,0.02}_{-0.03}$ & 6 \\
C50 & 53.9437 & -28.1163 & 6.54 & - & - \\
C51 & 53.9464 & -28.4051 & 5.64 & - & - \\
C52 & 53.9929 & -28.0131 & 2.67 & - & - \\
C53 & 54.0658 & -27.6901 & 6.59 & - & - \\
C54 & 54.0804 & -27.9724 & 5.65 & 1.55$^{+\,0.05}_{-0.05}$, 1.73$^{+\,0.02}_{-0.03}$ &  19, 7 \\
C55 & 54.1119 & -28.5234 & 7.25 & - & - \\
C56 & 54.1450 & -28.5939 & 3.34 & - & - \\
C57 & 54.2931 & -28.5204 & 2.56 & - & - \\
C58 & 54.2960 & -28.9641 & 3.76 & N/A & N/A \\
C59 & 54.3116 & -28.1648 & 4.66 & 1.73$^{+\,0.02}_{-0.03}$ & 4 \\
C60 & 54.3487 & -29.1917 & 3.31 & N/A & N/A \\
\end{tabular}
\label{TabGalsE}
\end{table*}

\begin{table*}
\centering
\contcaption{}
\begin{tabular}{wc{3cm} wc{1.2cm}  wc{1.2cm} wc{2cm} wc{2cm} wc{3cm}}
\hline
Group ID$^\textrm{\textit{a}}$ & RA$^\textrm{\textit{b}}$ & Dec$^\textrm{\textit{b}}$ & Size$^\textrm{\textit{c}}$ & Best $z_{phot}$ & Number Galaxies in \\
(AKA) & & & (arcmin$^2$) & estimate$^\textrm{\textit{d}}$ & Redshift Range$^\textrm{\textit{d}}$ \\
\hline \\ [-2ex]
C61 & 54.3492 & -28.5647 & 4.84 & - & - \\
C62 & 54.3699 & -28.7009 & 5.35 & - & - \\
C63 & 54.3959 & -27.9104  & 4.98 & - & - \\
C64 & 54.4928 & -28.3930 & 2.64 & - & - \\
C65 & 54.5010 & -28.8924 & 2.76 & N/A & N/A \\
\hline \\ [-2ex]
E1 & 7.2034 & -44.3381 & 2.47 & N/A & N/A\\
E2 & 7.2357 & -43.9173 & 13.0 & N/A & N/A\\
E3 & 7.4284 & -44.1157 & 2.43 & N/A & N/A\\
E4 & 7.4871 & -43.9178 & 4.37 & N/A & N/A\\
E5 & 7.5913 & -43.8762 & 4.52 & N/A & N/A\\
E6 & 7.8188 & -43.2838 & 3.64 & N/A & N/A\\
E7 & 7.8541 & -44.9474 & 2.43 & N/A & N/A\\
E8 & 8.1411 & -44.1827 & 4.95 & N/A & N/A\\
E9 & 8.3255 & -44.3937 & 3.01 & N/A & N/A\\
E10 & 8.3954 & -42.7188 & 2.45 & N/A & N/A\\
E12 & 8.4216 & -43.0649 & 2.98 & - & -\\
E13 & 8.5251 & -44.7918 & 6.14 & 1.43$^{+\,0.12}_{-0.08}$ & 47\\
E14 & 8.5726 & -45.1141 & 12.65 & - & -\\
E15 & 8.6012 & -45.0253 & 3.0 & - & -\\
E16 & 8.6191 & -45.1961 & 3.6 & N/A & N/A\\
E17 & 8.6431 & -44.1255 & 5.62 & 1.43$^{+\,0.02}_{-0.03}$ & 14 \\
E18 & 8.6964 & -45.2249 & 3.69 & N/A & N/A\\
E19 & 8.7452 & -43.6362 & 2.37 & - & -\\
E20 (SpARCS J0035-4312) & 8.9530 & -43.2096 & 5.91 & 1.18$^{+\,0.02}_{-0.03}$, 1.34$^{+\,0.06}_{-0.04}$ & 6, 32\\
E21 & 9.1957 & -45.4242 & 3.63 & N/A & N/A\\
E22 & 9.2864 & -42.4557 & 2.57 & N/A & N/A\\
E23 & 9.3699 & -45.1874 & 3.51 & N/A & N/A\\
E24 & 9.4261 & -42.6583 & 2.99 & N/A & N/A\\
E25 & 9.4878 & -44.8897 & 4.24 & 1.30$^{+\,0.05}_{-0.05}$, 1.54$^{+\,0.11}_{-0.14}$ & 18, 44\\
E26 & 9.4952 & -44.6427 & 8.45 & 1.43$^{+\,0.12}_{-0.13}$ & 95\\
E27 & 9.5237 & -44.2178 & 6.36 & 1.57$^{+\,0.13}_{-0.12}$ & 47\\
E28 & 9.5302 & -45.4546 & 2.54 & N/A & N/A\\
E29 & 9.5853 & -45.3231 & 12.4 & N/A & N/A\\
E30 & 9.6272 & -43.6191 & 6.92 & - & -\\
E31 & 9.6987 & -45.4063 & 3.66 & N/A & N/A\\
E32 & 9.7363 & -45.0858 & 3.66 & - & -\\
E33 & 9.7855 & -45.0291 & 2.48 & 1.39$^{+\,0.06}_{-0.04}$ & 16\\
E34 & 9.7963 & -42.9140 & 3.61 & 1.43$^{+\,0.12}_{-0.13}$ & 35\\
E35 & 9.8644 & -42.8679 & 2.43 & - & -\\
E36 & 9.8671 & -44.8111 & 4.15 & - & -\\
E37 & 9.9374 & -43.5112 & 2.38 & 1.48$^{+\,0.02}_{-0.03}$ & 6\\
E38 & 9.9444 & -43.1620 & 3.14 & - & -\\
E39 & 9.9863 & -43.1160 & 2.19 & - & -\\
E40 & 10.0292 & -43.8566 & 2.35 & - & -\\
E41 & 10.0410 & -44.3458 & 2.75 & - & -\\
E42 & 10.0425 & -44.4615 & 6.4 & - & -\\
E43 & 10.0450 & -44.5529 & 2.53 & 1.38$^{+\,0.07}_{-0.08}$ & 20\\
E44 & 10.1474 & -44.3490 & 2.36 & 1.52$^{+\,0.08}_{-0.07}$, 1.78$^{+\,0.02}_{-0.03}$ & 12, 2\\
E45 & 10.1782 & -43.8352 & 6.54 & - & -\\
E46 & 10.1918 & -44.2205 & 4.04 & 1.88$^{+\,0.07}_{-0.08}$ & 9\\
E47 & 10.2061 & -44.4126 & 2.48 & 1.43$^{+\,0.02}_{-0.03}$ & 6\\
E48 & 10.2843 & -43.9165 & 4.81 & 1.66$^{+\,0.04}_{-0.06}$, 1.83$^{+\,0.02}_{-0.03}$, 1.93$^{+\,0.02}_{-0.03}$ & 14, 4, 3\\
E49 & 10.3128 & -44.3738 & 6.17 & 1.34$^{+\,0.06}_{-0.04}$ & 26\\
E50 & 10.4522 & -44.4154 & 9.05 & - & -\\
E51 & 10.6005 & -43.0990 & 2.67 & N/A & N/A\\
E52 & 10.6160 & -43.9670 & 7.44 & N/A & N/A\\
E53 & 10.7067 & -42.6026 & 2.42 & N/A & N/A\\
E54 & 10.7255 & -44.3944 & 2.58 & N/A & N/A\\
E55 & 11.1364 & -43.5179 & 2.54 & N/A & N/A\\
E56 & 11.1560 & -43.2924 & 4.54 & N/A & N/A\\
E57 & 11.4203 & -44.0427 & 3.34 & N/A & N/A\\
E58 & 11.4875 & -43.3732 & 4.11 & N/A & N/A\\
\end{tabular}
\label{TabGalsC}
\end{table*}

\begin{table*}
\centering
\contcaption{}
\begin{tabular}{wc{3cm} wc{1.2cm}  wc{1.2cm} wc{2cm} wc{2cm} wc{3cm}}
\hline
Group ID$^\textrm{\textit{a}}$ & RA$^\textrm{\textit{b}}$ & Dec$^\textrm{\textit{b}}$ & Size$^\textrm{\textit{c}}$ & Best $z_{phot}$ & Number Galaxies in \\
(AKA) & & & (arcmin$^2$) & estimate$^\textrm{\textit{d}}$ & Redshift Range$^\textrm{\textit{d}}$ \\
\hline \\ [-2ex]
E59 & 11.5231 & -43.9767 & 2.36 & N/A & N/A\\
\hline \\ [-2ex]
X1$^{\ast}$ & 34.0953 & -5.0888 & 3.55 & - & -\\
X2$^{\ast}$ & 34.3090 & -4.5859 & 2.66 & 1.15$^{+\,0.05}_{-0.05}$ & 11\\
X3$^{\ast}$ & 34.3512 & -5.2810 & 3.22 & - & -\\
X4$^{\ast}$ & 34.3672 & -5.4229 & 3.86 & - & -\\
X5$^{\ast}$ & 34.4108 & -5.5328 & 3.78 & - & -\\
X6$^{\ast}$ & 34.4845 & -4.5380 & 2.80 & - & -\\
X7$^{\ast}$ & 34.4903 & -4.7543 & 6.33 & - & -\\
X8$^{\ast}$ & 34.5471 & -4.0601 & 3.23 & N/A & N/A\\
X9$^{\ast}$ (ClG J0218.3-0510) & 34.5877 & -5.1754 & 5.35 & - & -\\
X10$^{\ast}$ & 34.5940 & -4.5072 & 5.73 & - & -\\
X11$^{\ast}$ & 34.6235 & -4.6925 & 2.43 & - & -\\
X12$^{\ast}$ & 34.6734 & -5.3347 & 2.39 & - & -\\
X13$^{\ast}$ & 34.6873 & -5.2323 & 4.57 & - & -\\
X14$^{\ast}$ & 34.7555 & -3.5445 & 3.28 & N/A & N/A\\
X15$^{\ast}$ & 34.7982 & -4.7357 & 7.55 & - & -\\
X16$^{\ast}$ & 34.8032 & -6.1931 & 2.55 & N/A & N/A\\
X18$^{\ast}$ & 34.8309 & -5.2785 & 4.70 & - & -\\
X19$^{\ast}$ & 34.8441 & -4.4499 & 3.98 & 1.28$^{+\,0.02}_{-0.03}$, 1.48$^{+\,0.02}_{-0.03}$ & 8, 9\\
X20$^{\ast}$ & 34.8521 & -4.2207 & 4.39 & N/A & N/A\\
X21$^{\ast}$ & 34.8821 & -4.6303 & 14.95 & - & -\\
X22$^{\ast}$ & 34.9838 & -4.6538 & 5.65 & - & -\\
X23$^{\ast}$ & 35.0450 & -4.5589 & 4.17 & 1.65$^{+\,0.10}_{-0.10}$, 1.83$^{+\,0.02}_{-0.03}$, 1.98$^{+\,0.02}_{-0.03}$ & 21, 7, 8\\
X24$^{\ast}$ & 35.3648 & -4.1263 & 3.73 & - & -\\
X25$^{\ast}$ & 35.3780 & -5.5698 & 3.40 & - & -\\
X26$^{\ast}$ & 35.3897 & -4.1837 & 3.60 & - & -\\
X27$^{\ast}$ & 35.3978 & -4.6661 & 4.55 & - & -\\
X28$^{\ast}$ & 35.5678 & -4.3532 & 3.67 & - & -\\
X29$^{\ast}$ & 35.6106 & -4.2177 & 4.46 & 1.43$^{+\,0.02}_{-0.03}$, 1.53$^{+\,0.02}_{-0.03}$ & 17, 6\\
X30$^{\ast}$ & 35.6144 & -4.0216 & 5.99 & N/A & N/A\\
X31$^{\ast}$ & 35.6827 & -6.3192 & 4.41 & N/A & N/A\\
X32$^{\ast}$ & 35.7867 & -4.3803 & 11.58 & 1.42$^{+\,0.08}_{-0.07}$ & 80\\
X33$^{\ast}$ & 35.8040 & -4.4460 & 2.49 & 1.41$^{+\,0.14}_{-0.11}$ & 33\\
X34$^{\ast}$ & 35.8066 & -4.6453 & 4.88 & - & -\\
X35$^{\ast}$ & 35.8538 & -4.0661 & 2.06 & N/A & N/A\\
X36$^{\ast}$ & 35.8705 & -6.2554 & 2.46 & N/A & N/A\\
X37 & 36.0233 & -3.6699 & 3.07 & N/A & N/A\\
X38$^{\ast}$ & 36.0438 & -4.8454 & 3.94 & - & -\\
X39$^{\ast}$ & 36.0981 & -4.0080 & 2.39 & - & -\\
X40 & 36.1118 & -3.5421 & 3.88 & N/A & N/A\\
X41 (SpARCS J0224-0323) & 36.1257 & -3.4033 & 4.60 & N/A & N/A\\
X42$^{\ast}$ & 36.1385 & -5.4138 & 7.68 & - & -\\
X44$^{\ast}$ & 36.2382 & -4.2061 & 7.17 & - & -\\
X45$^{\ast}$ & 36.2826 & -4.6739 & 4.52 & - & -\\
X46$^{\ast}$ & 36.3119 & -4.7707 & 7.61 & - & -\\
X47 (SpARCS J0225-0355) & 36.4442 & -3.9330 & 7.27 & N/A & N/A\\
X48$^{\ast}$ & 36.5273 & -4.1293 & 2.70 & - & -\\
X50$^{\ast}$ & 36.5761 & -4.0365 & 2.47 & - & -\\
X51$^{\ast}$ & 36.6594 & -4.3120 & 4.12 & - & -\\
X52$^{\ast}$ (JKCS 041) & 36.6862 & -4.6956 & 3.05 & - & -\\
X53 & 36.6985 & -5.1840 & 12.31 & - & -\\
X54 & 36.7480 & -5.5329 & 3.22 & N/A & N/A\\
X55 & 36.7865 & -5.1939 & 2.41 & - & -\\
X56 & 36.8761 & -5.3278 & 4.99 & - & -\\
X58 (3XLSS J022734.1-041021) & 36.8954 & -4.1905 & 5.90 & - & -\\
X59 & 36.8993 & -4.1070 & 2.44 & - & -\\
X60$^{\ast}$ & 36.9727 & -4.5042 & 3.84 & - & -\\
X61 & 36.9988 & -5.0176 & 2.59 & - & -\\
X62 (3XLSS J022806.4-044803) & 37.0301 & -4.8026 & 4.60 & - & -\\
X63 & 37.1094 & -5.1443 & 2.36 & - & -\\
X64 & 37.1553 & -4.6027 & 3.35 & - & -\\
X65 & 37.1621 & -4.9243 & 2.48 & 1.30$^{+\,0.15}_{-0.15}$ & 36 \\ [0.5ex]
\hline
\end{tabular}
\end{table*}

\begin{table*}
\centering
\contcaption{(likely low-redshift contaminants)}
\begin{tabular}{wc{3cm} wc{1.2cm} wc{1.2cm} wc{2cm} wc{2cm} wc{3cm}}
\hline
Group ID$^\textrm{\textit{a}}$ & RA$^\textrm{\textit{b}}$ & Dec$^\textrm{\textit{b}}$ & Size$^\textrm{\textit{c}}$ & Best $z_{phot}$ & Number Galaxies in \\
(AKA) & & & (arcmin$^2$) & estimate$^\textrm{\textit{d}}$ & Redshift Range$^\textrm{\textit{d}}$ \\
\hline \\ [-2ex]
E11 (XClass 534) & 8.4091 & -43.2981 & 2.54 & - & -\\ 
X17$^{\ast}$ & 34.8186 & -5.1750 & 3.28 & 0.38$^{+\,0.02}_{-0.03}$ & 7\\ 
X43$^{\ast}$ & 36.1864 & -4.9369 & 8.80 & 0.34$^{+\,0.06}_{-0.04}$ & 21\\ 
X49$^{\ast}$ (XClass 20372) & 36.5684 & -4.9527 & 6.44 & - & -\\ 
X57$^{\ast}$ (XClass 476) & 36.8791 & -4.5453 & 3.66 & 0.28$^{+\,0.02}_{-0.03}$ & 4\\ [0.5ex]
 \hline \\ 
\end{tabular}
 \small
 \begin{flushleft} \textit{Notes}. $^\textrm{\textit{a}}$Groups with IDs beginning with C are located in the CDFS, E in ELAIS S1 and X in XMMLSS. \\ 
 $^\textrm{\textit{b}}$Defined as the mean position of the selected red IRAC galaxies. \\
 $^\textrm{\textit{c}}$Defined as the area enclosed within the boundary of the associated red IRAC galaxies. \\
 $^\textrm{\textit{d}}$N/A if group does not fall within footprint of photo-$z$ catalogues, - if no redshift peak can be identified.
 \end{flushleft}
\end{table*}

\bsp	
\label{lastpage}
\end{document}